\documentclass[manusript]{aastex7}
\usepackage{stix2}
\usepackage[T1]{fontenc}
\usepackage[]{graphicx,tabularx,ctable,color,colortbl,amsmath,amsfonts,float,bm}
\usepackage{anyfontsize}

\shortauthors{}

\begin{document}

\title{Frequency-time-resolved Imaging Spectroscopy of Fine Structures in a Solar Radio Noise Storm}

\author[0000-0003-1967-5078]{Daniel L. Clarkson}
\affiliation{School of Physics \& Astronomy, University of Glasgow, Glasgow, G12 8QQ, UK}
\email{daniel.clarkson@glasgow.ac.uk}  

\author[0000-0002-8078-0902]{Eduard P. Kontar}
\affiliation{School of Physics \& Astronomy, University of Glasgow, Glasgow, G12 8QQ, UK}
\email{eduard.kontar@glasgow.ac.uk}  

\begin{abstract}
    Solar radio noise storms are common phenomena, 
    composed of broadband continuum emission 
    embedded with diverse fine structures, yet their unusually compact 
    apparent sizes remain unexplained. 
    We present frequency-time-resolved imaging spectroscopy 
    of a near-disk-center noise storm observed by LOFAR 
    between 30--40 MHz, together with anisotropic radio-wave scattering 
    simulations. The continuum forms a bright, spatially compact source 
    that drifts across the solar disk over tens of minutes. 
    Across the band, the measured apparent major axis 
    is $\sim8.0\arcmin$ to $\sim4.3\arcmin$ between 31.3 and 38.4~MHz, 
    less than half the size  of typical type III burst sources at comparable 
    frequencies. Embedded type I bursts, S-bursts, and spikes 
    exhibit a range of dynamic spectra appearances, 
    yet share nearly identical apparent sizes within uncertainties, 
    suggesting a common size-determining mechanism.
    Using anisotropic scattering simulations, we show that compact 
    apparent source sizes naturally arise for emission embedded within 
    closed magnetic field structures, where anisotropic turbulence 
    directs radiation away from the observer's line of sight. 
    Additional modifications arise from enhanced coronal densities, 
    steeper density gradients, reduced turbulence levels, 
    and strong fluctuation anisotropy, but these exert secondary influence. 
    Our results provide a unified explanation for the similar apparent sizes 
    of diverse fine structures in noise storms, and demonstrate 
    that the compactness of type I storm sources is governed primarily 
    by the large-scale coronal environment rather than intrinsic differences 
    in emission processes, where the magnetic topology plays a crucial role 
    in determining the observed source size.
\end{abstract}


\section{Introduction}

    Solar radio noise storms are a common, yet puzzling type of radio emission. They are strongly circular polarized and composed of a broadband continuum overlaid with numerous short-lived fine-structures called type I bursts \citep[see][as a review]{1985srph.book..415K}. 
    Noise storms can last for hours to days and are often associated with large, complex sunspot groups \citep{1951AuSRA...4..508P, 1970SoPh...11..456K, 1985SoPh...96..381S,1985srph.book..415K}, with the storm source observed to lie within coronal loops that can span two active regions \citep{1970SoPh...11..456K, 1985srph.book..415K,2021ApJ...920...11M}.  Individual type I bursts often appear in clusters, are typically sub-second in duration, and have bandwidth ratios between 2\%--3\% \citep[e.g.][]{1970A&A.....5..372E,1972SoPh...24..215C,1976SoPh...48..321D,1982SoPh...77..231K,1985srph.book..415K,1989ApJ...342..594H,1991SoPh..132..155T,2004ApJ...605..948S,2025ApJ...985..257Y}. Type I bursts are distinct from other types of fine structure such as S-bursts and spikes. At decameter frequencies, S-bursts appear in groups and have instantaneous bandwidths of in the hundreds of kilohertz range \citep{1982SoPh...78..253M, 2010SoPh..264..103M} covering total bandwidths of a few megahertz, clear negative frequency drift rates of a few megahertz per second, and durations $\sim0.5$~s \cite[e.g.][]{2010SoPh..264..103M, 2015A&A...580A..65M}. Decameter spikes are narrower with bandwidth ratios between 0.1\%--0.6\% and durations $\lesssim1$~s \citep{2014SoPh..289.1701M, 2023ApJ...946...33C}, and typically have absolute frequency drift rates near zero up to tens of kilohertz per second at 30--50 MHz \citep[e.g.][]{Clarkson_2021, 2023ApJ...946...33C}.
    
    The emission mechanism of type I bursts remains uncertain, with numerous proposed models. Their high brightness temperatures and polarization, along with a lack of observable harmonic component, suggest fundamental plasma emission due to the coalescence or decay of Langmuir waves and low-frequency turbulence (e.g. ion-sound waves). \citet{1980SoPh...67..357M} suggests that the required generation of Langmuir waves is associated with a loss-cone distribution of trapped electrons in closed magnetic structures, while the low-frequency turbulence is likely present where the corona is being heated, particularly over active regions. However, this scenario does not explain why no harmonic emission is observed. \citet{2017SoPh..292..117L} propose that in the low beta plasma of active regions, imbalanced turbulence of inertial Alfvén waves can facilitate the growth of Langmuir waves, leading to fundamental radio emission. In this model, no harmonic emission is generated since Landau damping remains strong for the negative wavenumber Langmuir waves. \citet{2014A&A...562A..57R} show that negative wavenumber Langmuir waves can be suppressed due to strong turbulence in coronal loops when the diffusive timescale becomes shorter than the quasilinear timescale. Other suggestions involve moving magnetic features, including small-scale magnetic reconnection \citep{2000SoPh..193..227B, 2017SoPh..292...82L}.

    \begin{figure}[htb!]
        \centering
        \includegraphics[width=0.85\textwidth]{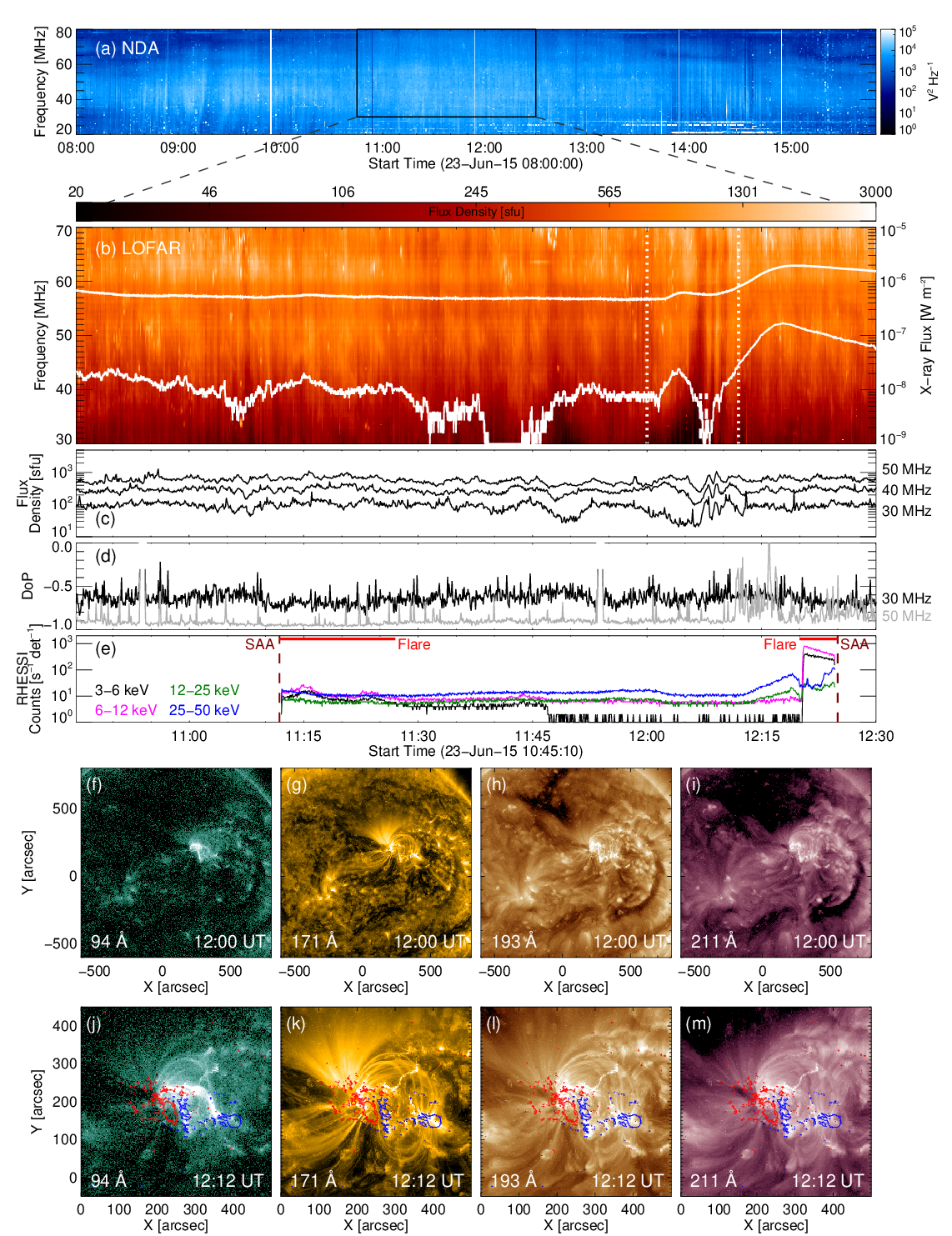}
        \caption{Radio, X-ray, and EUV observations. \textbf{(a)} NDA
        observed 8 hour dynamic spectrum.
        Time (UT) is shown in 30-minute intervals. The black box shows the
        times and frequencies observed by LOFAR. \textbf{(b)} LOFAR
        observed dynamic spectrum across 105 minutes between
        30--80~MHz, overlaid with the GOES X-ray flux (white curves)
        between 0.5--4~\AA\,(lower) and 1--8~\AA\, (upper). Time (UT) is shown in 15-minute intervals.
        The vertical white dashed lines show the
        times and frequencies imaged in Figure \ref{fig:clean_im_freq}
        (see also Figure \ref{fig:ds_zoom_flare}). \textbf{(c)} LOFAR time
        profiles at 30, 40, and 50~MHz, reduced to a time resolution
        of 5.2~s.
        \textbf{(d)} Degree of polarization (DoP) observed by the
        Nan\c{c}ay Decameter Array (NDA) reduced to 5~s time
        resolution. \textbf{(e)} RHESSI count rates in various energy
        bands. The red bars indicate times of flaring activity, and
        the brown dashed lines denote times where the spacecraft
        leaves and enters the South Atlantic Anomaly (SAA). SDO/AIA
        images at different wavelengths are shown near 11:53 UT
        \textbf{(f-i)} and 12:12 UT \textbf{(j-m)}. The red and blue
        contours represent the HMI magnetogram at 450 and $-450$~G,
        respectively.}
        \label{fig:ds_2hr_goes_aia}
    \end{figure}
    
    Imaging analysis \cite[e.g.][]{1987ApJ...319..514L,1973PASA....2..211D, 2015A&A...576A.136M, 2024ApJ...975..122M, 2025SoPh..300..109M}, indicates that type I storm source sizes are smaller than type III source sizes 
    \cite[][]{1970SoPh...11..456K}.
    This is puzzling, given that radio-wave scattering, which dominates the apparent source size, does not differentiate between burst types \citep[][]{2019ApJ...884..122K}. Moreover, studies investigating the spatial variation of frequency-time structures in noise storms are virtually non-existent, as most observations are conducted at a single frequency. At decimeter frequencies, \citet{1987ApJ...319..514L} observed sizes of $40^{\mathit{\prime}\mathit{\prime}}$ for four compact sources within an elongated source, using the VLA with $9^{\mathit{\prime}\mathit{\prime}}$ angular resolution at 328 MHz. At 333 MHz, \cite{1992SoPh..141..165Z} 
    used $4^{\mathit{\prime}\mathit{\prime}}$ resolution VLA imaging 
    and found source sizes no smaller than $30^{\mathit{\prime}\mathit{\prime}}$, with typical sizes between 40 and 90 arcseconds. More recently, 
    \cite{2015A&A...576A.136M} combined Nan\c{c}ay Radioheliograph and Giant Meterwave Radio Telescope (GMRT) visibilities to achieve $20\arcsec$ resolution 
    at 236 and 327 MHz, finding compact cores with minimum sizes between 31\arcsec--35\arcsec. 
    Using the upgraded GMRT, \cite{2024ApJ...975..122M} observed 
    source sizes between approximately 9\arcsec--20\arcsec\, in the 221--251~MHz range.
    
    The small sizes of noise storms are of particular interest when
    considering the minimum observable sizes of radio sources limited
    by radio-wave scattering. \cite{2019ApJ...884..122K, 2023ApJ...956..112K} 
    use an anisotropic turbulence model to replicate radio burst characteristics
    and calculate coronal heating rate \citep{2025ApJ...991L..57K}, 
    finding the required wavenumber density anisotropy to be typically between
    $\alpha=q_\parallel/q_\perp=(0.25-0.4)$. Strong anisotropy ($\alpha\ll1$) 
    was found to have a somewhat weak affect on source
    size, although this effect was more pronounced towards the limb
    owing to the direction of the anisotropy axis compared to the line of sight. Both \cite{2020ApJ...898...94K} and \cite{Clarkson_2021,2023ApJ...946...33C} consider values of $\alpha\simeq0.1$ to explain the small sizes of drift-pair bursts, and radio spikes. In this context, \cite{2023ApJ...956..112K} use spike measurements (the shortest solar radio bursts in the decameter range) 
    to infer the minimum observable apparent source size at a given frequency, limited by scatter-broadening, and require a scaling factor of 0.5--2 on
    their turbulence model to account for the spread in observational data, including both solar radio bursts and extrasolar radio sources. 
    In the context of anisotropic scattering, the viewing
    angle relative to the magnetic field in the scattering region also affects the apparent source size. The observer sees a reduced
    spatial extent if the magnetic field direction within the region
    of strong scattering is directed away from the observer since the
    bulk of radiation escape is away from their line of sight. Indeed,
    simulations of radio point sources in dipolar magnetic field
    structures show remarkably small source sizes
    \citep{2025ApJ...978...73C} compared to similar simulations for
    sources within open fields \citep{2019ApJ...884..122K}. Numerous
    type I noise storm events are observed to be elongated structures
    \cite[e.g.][]{1987ApJ...319..514L, 2015A&A...576A.136M,2025SoPh..300..109M}, and their narrow minor axes have been used
    as evidence that the level of scattering is overestimated
    \cite[e.g.][]{2015A&A...576A.136M}. Moreover, previous
    observations of type I nose storms have placed their source
    heights farther from the Sun than expected, indicating over-dense
    regions \citep{1976SoPh...50..437S, 2015A&A...576A.136M}.
    Simulations by \cite{2025SoPh..300..109M} between (150--500)~MHz
    show that over-dense regions of the corona can lead to a reduction
    in source sizes by $3.5\times$ compared to simulations with
    typical background density levels; however, the smallest simulated
    size was $\sim50\arcsec$, more than double the
    smallest observed size in their study at $\sim20\arcsec$. We
    further note that the simulations did not include a non-radial
    magnetic field structure.
    
    In this study, we use LOFAR observations between 30--40~MHz to investigate the characteristics of embedded fine structures in a noise storm (type I bursts, S-bursts, and spikes) at sub-second scales. We combine imaging observations of the continuum and type I bursts with radio-wave propagation simulations to investigate the compact sizes of a type I noise storm for a source close to the solar disk center. We show that compact source sizes can result from anisotropic radio-wave scattering for sources located within closed magnetic fields, as well as from over-dense regions that modify the source height and density gradient, with additional contributions from stronger anisotropy and reduced turbulence.
      
    Section \ref{sec:event} outlines the observed event. Sections \ref{sec:data_methods} and \ref{sec:lofar_results} 
    describe the observational methods and findings on the source position and size, while section \ref{sec:sim_results} outlines the simulation approach and the plasma conditions that produce compact source sizes. Sections \ref{sec:discussion} and \ref{sec:conclusion} present the discussion and conclusion.

\section{Overview of the Event}\label{sec:event}

    \begin{figure}[b!]
        \centering
        \includegraphics[width=0.9\textwidth]{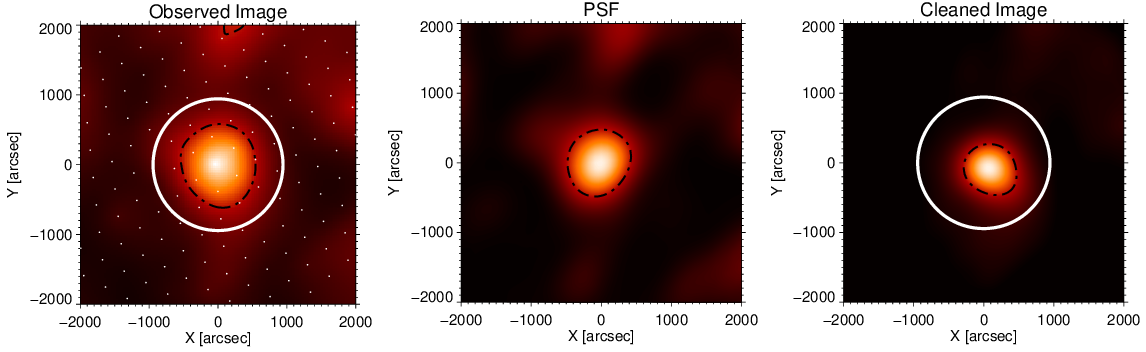}
        \caption{\textbf{(a)} Observed ``dirty'' LOFAR intensity map
        at 32.5~MHz at 12:13:35 UT. The white dots represent the
        phased array beam locations. \textbf{(b)} LOFAR effective PSF
        using 24 combined observations from Tau A at zenith and
        rotated to the solar position \citep{2022ApJ...925..140G}.
        \textbf{(c)} Resulting LOFAR cleaned map. The white circles
        represent the solar disk, and the dot-dashed contours outline
        the intensity at the 50\% levels.}
        \label{fig:clean_eg}
    \end{figure}
    
    We report on a type I noise storm observed on 23 June 2015 when
    the active region AR12371 west from the solar disk center.
    The storm was observed throughout the daylight hours by the
    \emph{Nan\c{c}ay Decameter Array}
    \citep[NDA;][]{2000GMS...119..321L} between 20--80 MHz, and by the
    \emph{LOw Frequency ARray} \citep[LOFAR;][]{2013A&A...556A...2V}
    between 30--80~MHz for almost two hours from 10:45 to 12:30 UT (Figure
    \ref{fig:ds_2hr_goes_aia}).
    The LOFAR dynamic spectrum shows broadband continuum emission
    across the observed frequency range (Figure
    \ref{fig:ds_2hr_goes_aia}b). The flux density is brighter between
    50--70~MHz than at 30~MHz by almost an order of magnitude,
    reaching up to $\sim10^3$~sfu at 50~MHz and $\sim10^2$~sfu at
    30~MHz (Figure \ref{fig:ds_2hr_goes_aia}c). The continuum is
    overlain by numerous brighter type I bursts (e.g. Figure
    \ref{fig:cont_typeI_ds_sizes}a). Additional fine structures are
    present across certain time intervals such as S-bursts, and spikes
    (Figure \ref{fig:sbursts_spikes_ds}), distinguished from type I
    bursts by their total and instantaneous bandwidths, and their frequency drift rates.
    During the LOFAR observing period, 
    there were two small flares detected by \emph{Reuven Ramaty High Energy Solar Spectroscopic Imager}
    \citep[RHESSI;][]{2002SoPh..210....3L} (Figure \ref{fig:ds_2hr_goes_aia}c), and a long period without visible flaring activity. The flare around 12:20 UT is also well seen as a C-class flare 
    in soft X-ray (SXR) flux observed by the \emph{Geostationary Operational Environmental Satellite} (GOES) in Figure \ref{fig:ds_2hr_goes_aia}b. 
    Polarization measurements from NDA show strong circular 
    polarization 70\%--100\% 
    for the duration of the observations with higher
    polarization degree at 50~MHz compared to 30~MHz 
    (Figure \ref{fig:ds_2hr_goes_aia}d). 
    AIA imaging highlights three magnetically connected sunspot 
    groups near the disk center shown in Figure \ref{fig:ds_2hr_goes_aia}(f-i). The zoomed-in panels (j–m) show an active region 
    with a leading sunspot of negative polarity 
    and a trailing group that exhibits both positive and 
    negative polarities (see also
    \cite{2017ApJ...845...59V,2020SoPh..295...29M,2021Ge&Ae..61S..24P}
    for detailed magnetogram analyses of this region over successive
    days). The trailing group is larger than the leading spot, with
    the positive-polarity region being more extended and complex than
    the negative component, suggesting that it is the dominant
    structure. This configuration, in which the trailing region is
    dominant, has been reported in a substantial fraction of cases
    (around one third; \citealp{1985SoPh...96..381S}) and implies that
    the noise storm emission is left-hand circularly polarized
    \cite[e.g.][]{1973PASA....2..211D,1985srph.book..415K}. 
    The active region shows a complex magnetic loop arrangement 
    with a fan-like structure at its northern edge. 
    The loops present frequent transient brightening, 
    and emission of plasma jets occur throughout the observing period. 
    The second flaring episode at
    12:20 UT is preceded at 12:12 UT by loop brightening that bridges
    regions of opposite magnetic polarity.
    
\section{Observational Data Processing and Analysis
Methods}\label{sec:data_methods}

    The LOFAR imaging observations were performed in tied-array beam
    forming mode using 24-core Low Band Antenna (LBA) stations in the
    outer configuration \citep{2013A&A...556A...2V} with a temporal
    and spectral resolution of 10~ms and 12.2~kHz, respectively. 
    The tied-array beams were composed of 169 individual and overlapping beams pointing across the solar disk out to $\sim2$~R$_\odot$ with $\sim0.1\arcdeg$ spacing between each beam center. Two additional
    beams were pointing at Tau A, used as a calibrator
    \citep{2017NatCo...8.1515K}. We focus on imaging frequencies
    between 30--40~MHz where the sidelobe emission is minimal
    \cite[e.g.][]{2019ApJ...873...48G}, for which the baseline $D$ of
    $\sim3.5$~km provides a nominal beam size 
    of $\theta\approx\lambda/D\simeq 10 \prime$ around 30~MHz.

    \begin{figure}[b!]
        \centering
        \includegraphics[width=0.9\textwidth]{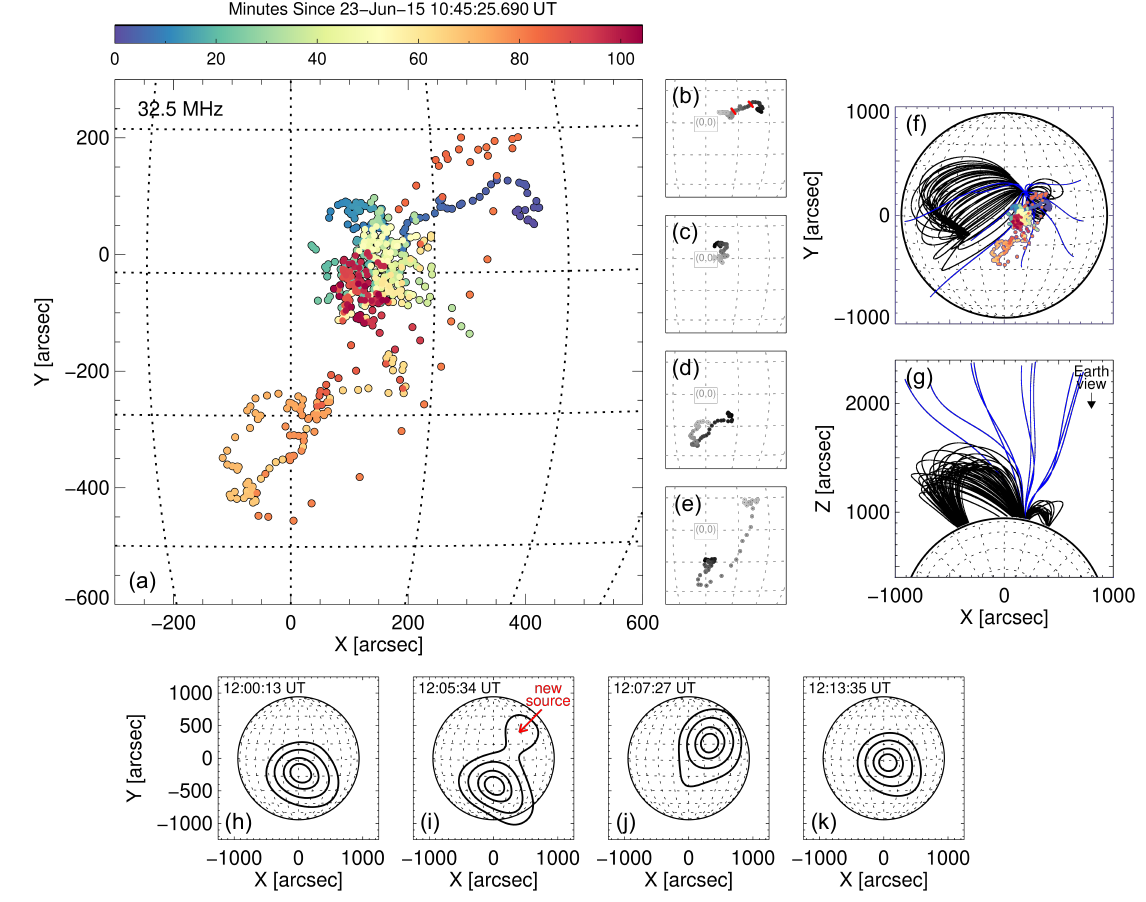}
        \caption{Source positions over time. (a) Centroid position of the cleaned LOFAR images derived from 2D elliptical Gaussian fitting over time at 32.5~MHz. Each centroid is 10 seconds apart. Panels (b-e) show four selected 10 minute intervals where the centroids show pronounced movement in a particular direction over time from black to light grey. The point (0,0) denotes the center of the solar disk. The red dashes in panel (b) bound a sky-plane distance of $145\arcsec$ linear motion across 160~s. Panels (f, g) show a potential-field source-surface (PFSS) extrapolation for open (blue) and closed (black) field lines, as viewed from Earth (upper), and rotated $-90$\arcdeg\, in latitude (lower). Panels (h-k) show radio lobe contours over time at 30\%, 50\%, 75\%, and 90\% levels, and highlight the appearance of a new source near 12:05 UT that causes the centroid to shift towards the flaring region in panel (e).}
        \label{fig:source_pos_time}
    \end{figure}
    
    Following \cite{2017NatCo...8.1515K,2022ApJ...925..140G}, 
    the resulting ``dirty map'' (e.g. Figure \ref{fig:clean_eg}a) is passed through a CLEAN algorithm to minimize sidelobe emission, where an effective point-spread function (PSF) is determined from 24 combined observations of Tau
    A at zenith and translated to the solar position with a zenith
    angle of $\sim31\arcdeg$ throughout the observation (Figure
    \ref{fig:clean_eg}b). Additionally, a frequency-dependent
    correction for average ionospheric refraction is applied to each
    radio map, which shifts the clean image along the zenith direction (see equation 6 of \cite{2022ApJ...925..140G}). 
    The observed sizes and positions of the radio bursts are determined by fitting a 2D elliptical Gaussian to the clean maps $I(x,y)$ (e.g. Figure \ref{fig:clean_eg}c) as
    \begin{equation}\label{eq:2dgauss}
        I(x, y) = I_0\exp\left(-\frac{x\prime^2}{2\sigma_x^2} - \frac{y\prime^2}{2\sigma_y^2}\right),
    \end{equation}
    where $x^\prime = \left(x - x_s\right)\cos{T} - \left(y -
    y_s\right)\sin{T}$ and $y^\prime = \left(x - x_s\right)\sin{T} -
    \left(y - y_s\right)\cos{T}$, $I_0$ is the peak flux density
    amplitude, $x$ and $y$ are the sky-plane map coordinates,
    $\sigma_x$ and $\sigma_y$ are the standard deviations from which
    the FWHM major $S_\mathrm{maj}$ and minor $S_\mathrm{min}$ axes,
    related to the Gaussian widths by $S=2\sqrt{2\ln{2}}\,\sigma$. The source centroid locations in the sky-plane are $x_s$ and $y_s$,
    and $T$ is the ellipse rotation angle clockwise from the
    \textit{x}-axis. The FWHM source area is given as $A=\pi/4\cdot
    S_\mathrm{maj}\cdot S_\mathrm{min}$. The uncertainties on the
    centroid position $(\delta{x_c}, \delta{y_c})$ as given by
    \cite{1997PASP..109..166C}:
    \begin{equation}
        \begin{split}
            \delta{x_c} &= \sqrt{\frac{2}{\pi}\frac{\sigma_x}{\sigma_y}}\frac{\delta{S}}{S_0}\theta\\
            \delta{y_c} &= \sqrt{\frac{2}{\pi}\frac{\sigma_y}{\sigma_x}}\frac{\delta{S}}{S_0}\theta,
        \end{split}        
    \end{equation}
    and $\delta{S}$ is the flux uncertainty determined as the average
    flux value from half the faintest beams which form a given image
    under the assumption that in at least half the field-of-view, the
    real signal is not greater than the noise
    \citep{2022ApJ...925..140G}. The fractional uncertainty in the
    source sizes are given as
    \begin{equation}
            \frac{\delta\sigma_x}{\sigma_x} = \frac{\delta\sigma_y}{\sigma_y} = \sqrt{\frac{2}{\pi}}\frac{1}{\sqrt{\sigma_x\sigma_y}}\frac{\delta{S}}{S_0}h,
    \end{equation}
    leading to uncertainty on the area as
    \begin{equation}
        \frac{\delta{A}}{A} = \sqrt{ \left(\frac{\delta\sigma_x}{\sigma_x}\right)^2 + \left(\frac{\delta\sigma_y}{\sigma_y}\right)^2 } = \sqrt{2}\frac{\delta\sigma_x}{\sigma_x}.
    \end{equation}
        
    Where the fine structures are sufficiently bright above the
    continuum level, their frequency drift rates,
    $\mathrm{d}f/\mathrm{d}t$, are determined following the analysis
    approach used in \cite{2018SoPh..293..115S, Clarkson_2021}. For
    each time, a Gaussian function of the form
    $I(f)=I_0\exp{(-(f-f_0)^2/(2(\Delta{f})^2)) +
    I_\mathrm{cont}+I^\prime_\mathrm{cont} f}$, is fitted to the frequency-flux profile
    $I(f)$ and frequency of the maximum flux tracks the evolution of
    the fine structure drift. Here, $I_0$ is the peak flux, $f_0$ is
    the frequency of maximum emission, $I_\mathrm{cont}$ is the
    background level, and the final term accounts for any linear
    change in the continuum flux density with frequency.
    \begin{figure}[htb!]
        \centering
        \includegraphics[width=.9\textwidth]{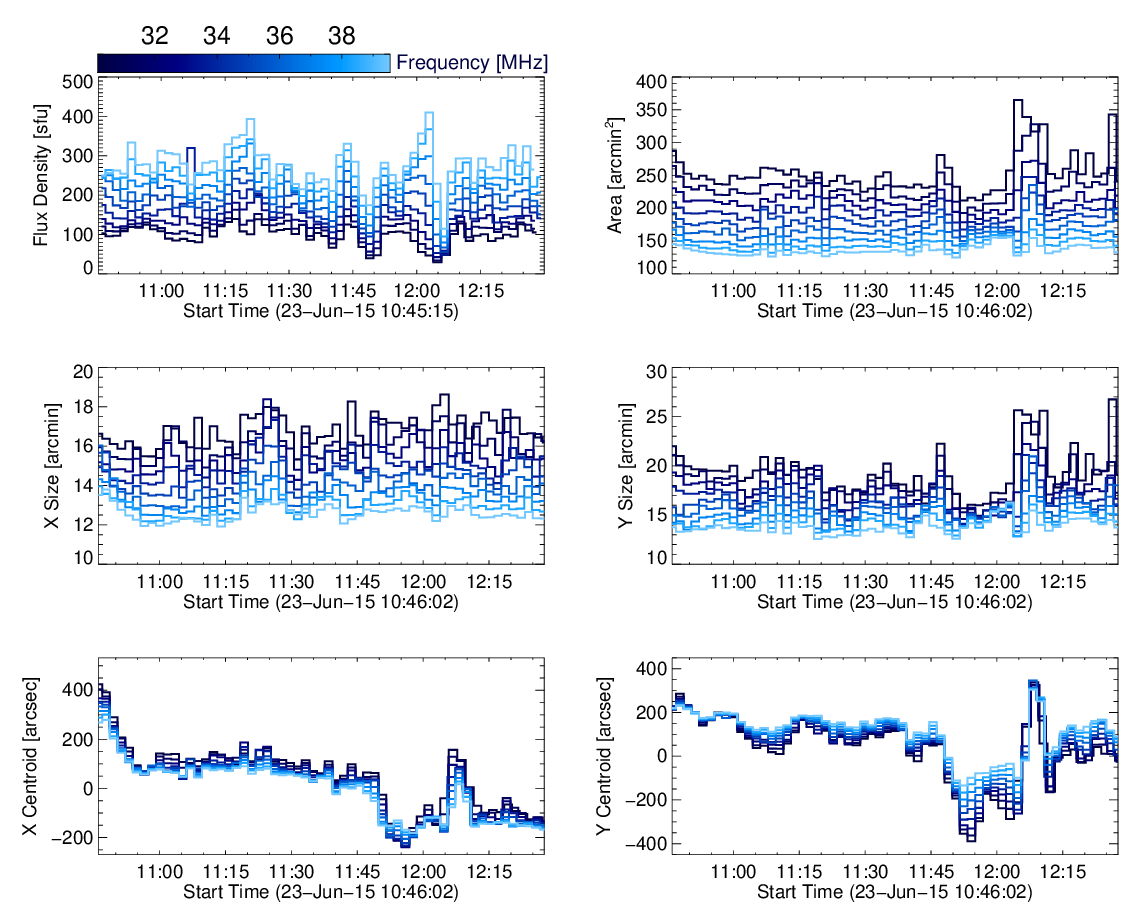}
        \caption{Noise storm characteristics over 105~minutes averaged across 1~MHz frequency channels between 30--40~MHz at 100~s time intervals. Time (UT) is shown in 15-minute intervals. (a) Flux Density. (b) FWHM Area. (c) FWHM size in the x-direction. (d) FWHM size in the y-direction. (e) Centroid position in the x-direction. (f) Centroid position in the y-direction.}
        \label{fig:2hr_char}
    \end{figure}
    The uncertainty on the observed flux density is given as
    $\delta{I}=\sqrt{(0.12I)^2 + I_\mathrm{cont}^2}$, which combines a
    12\% uncertainty in the flux density with the typical continuum
    level in frequency space surrounding the fine structure, in
    quadrature. The collated peak frequencies and times are fit with a
    linear model of the form $f(t)=(\mathrm{d}f/\mathrm{d}t)t+f_c$
    using the $1\sigma$ uncertainties in the position of the peak
    frequencies returned form the Gaussian fitting. From the linear
    fit, the drift rate is determined. For the S-bursts, where the
    contrast is low compared to the continuum, the drift rate is
    estimated by manually tracing the feature with a straight line in
    time and frequency. The uncertainty is then determined by defining
    the steepest and shallowest plausible slopes that still follow the
    feature. The adopted uncertainty is taken as half the difference
    between these limiting drift rates.

    \begin{figure}[htb!]
        \centering
        \includegraphics[width=0.6\textwidth]{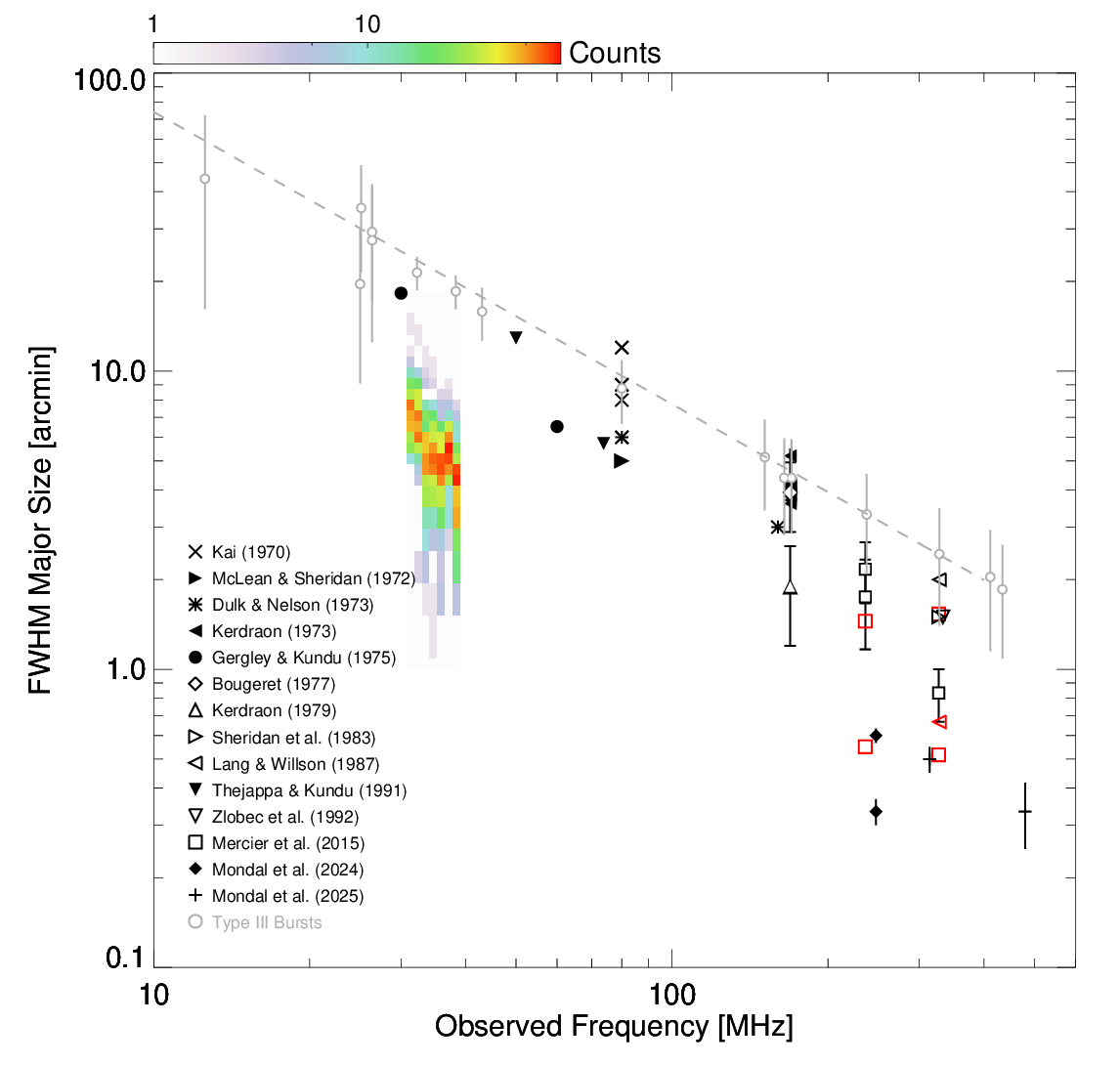}
        \caption{Noise storm and type I burst FWHM major sizes against
        frequency. The histogram data shows 2660 measurements of the
        cleaned and beam-corrected (as $S=\sqrt{S_\mathrm{clean}^2 -
        S_\mathrm{psf}^2}$) LOFAR observations throughout 105 minutes across 1~MHz averaged frequency intervals from 2D elliptical
        Gaussian fits (see Figure \ref{fig:clean_eg} for an example of the images). The binning of
        the size axis is 0.5~arcmin. The black data shows the major
        sizes from \cite{1970SoPh...11..456K, 1972SoPh...26..176M,
        1973PASA....2..211D, 1973A&A....27..361K, 1975SoPh...41..163G,
        1977A&A....60..131B, 1979A&A....71..266K, 1983SoPh...83..167S,
        1987ApJ...319..514L, 1991SoPh..132..155T, 1992SoPh..141..165Z,
        2015A&A...576A.136M, 2024ApJ...975..122M,
        2025SoPh..300..109M}. Data with both black and red symbols
        shows the size of the core (red) and halo (black) if
        specified. The data for \cite{1979A&A....71..266K} shows the
        average and standard deviation from a distribution of measured
        sizes (their Figure 1) which also includes some complex
        (multiple) sources. Data for \cite{1975SoPh...41..163G} is
        from 1D scans and beam corrected according to their equation 7. Half-power data from
        \cite{1991SoPh..132..155T} are corrected for their quoted beam
        sizes. Data from \cite{1977A&A....60..131B} is the average and
        standard deviation from the beam-corrected data in
        their Figure 4. The measurement level of the source diameters is not mentioned,
        so we consider these values an upper limit. The gray
        data show the type III burst sizes compiled in
        \cite{2019ApJ...884..122K} and corresponding fit (gray dashed
        line).}
        \label{fig:typeI_sizes}
    \end{figure}
        
\section{LOFAR Dynamic Spectrum and Imaging
Analysis}\label{sec:lofar_results}

    \subsection{Evolution Across Two Hours}\label{sec:evolution_2hr}

        Imaging of the noise storm at a fixed frequency reveals that
        the radio source drifts across the solar disk over time. At
        32.5~MHz (Figure \ref{fig:source_pos_time}), the source
        centroid starts near the active region and shifts south-east
        over approximately 80~minutes before returning above the
        active region, covering a distance of approximately
        $800\arcsec$ (0.8~R$_\odot$). The apparent jump-like 
        motion north-west, highlighted in panel (e), is likely due to the emergence of a
        new radio source (panel (i)), causing the centroid to represent the
        combined position of two sources for several minutes until the  earlier source is no longer detected.
        This new source appears near the timing of the flaring episode,
        and is likely related to the injection of a new population of electrons
        to the loop system from a reconfiguring magnetic field.
        Figure \ref{fig:2hr_char} shows the noise storm characteristics
        across the almost two hour observation, averaged over 1~MHz frequency bandwidths and $\sim100$~s time intervals. On scales for tens of minutes to hours, the flux density, sizes, and areas show
        little variation, with the main change being 
        in the source position.

        \begin{figure}[htb!]
            \centering\includegraphics[width=0.5\textwidth]{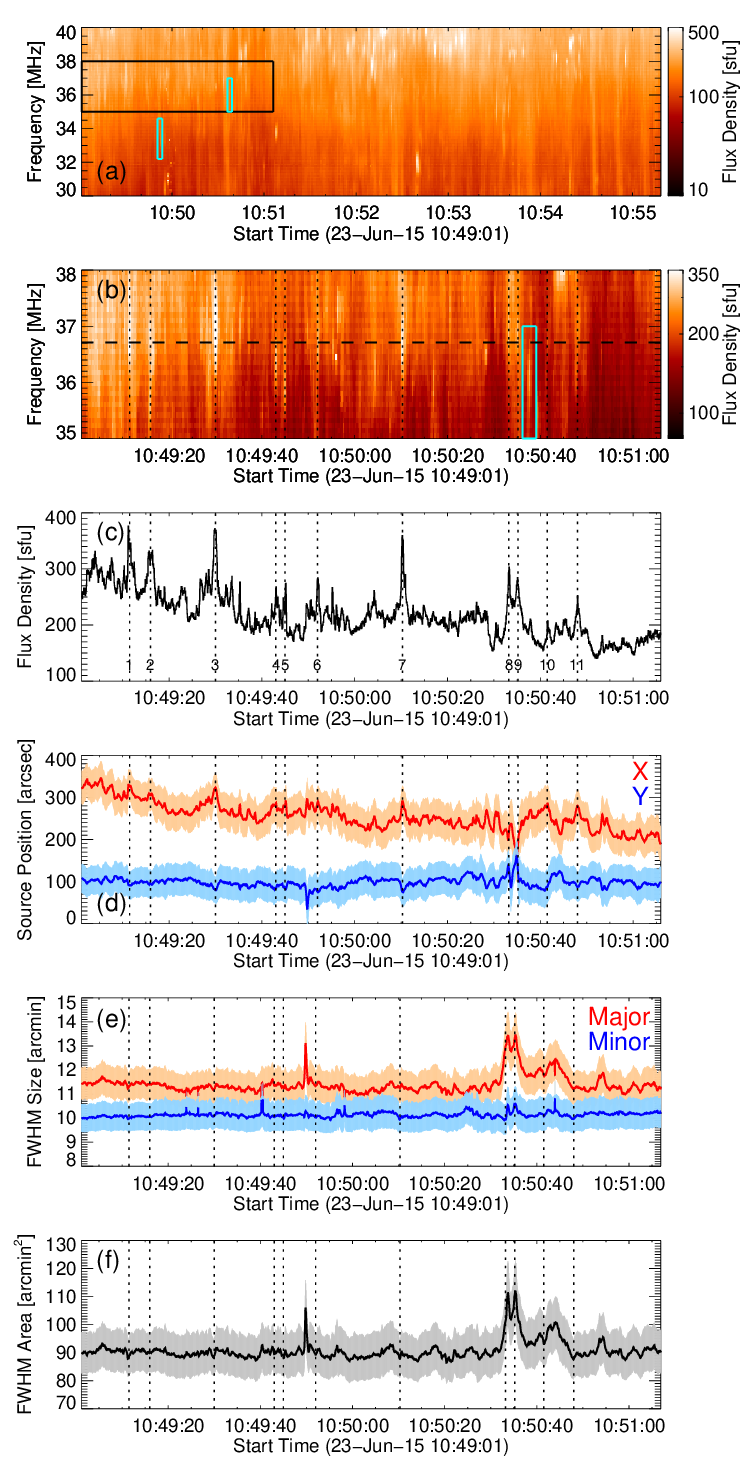}
            \caption{Characteristics of continuum emission and
            type I bursts across across approximately one minute.
            (a) Dynamic spectrum showing continuum emission
            overlaid with type I bursts over six minutes for context.
            Time (UT) is shown in 1-minute intervals.
            The black box highlights the zoomed in region in panel
            (b). The light blue boxes highlight the individual type I
            bursts shown in Figure \ref{fig:typeI_dfdt_size}. (b)
            Zoomed in dynamic spectra. The black vertical dotted lines
            highlight examples of bright type I bursts. The black
            horizontal dashed line shows the frequency channel at
            36.73~MHz corresponding to panels (c)-(f). (c) Time
            profile. (d) Source position in the sky-plane for X (red)
            and Y (blue) coordinates. (e) FWHM cleaned major and minor sizes over time.
            The orange and light blue color shows the uncertainty on
            each data point. Data points in all panels are separated
            by 50~ms. (f) FWHM cleaned area over time.
            The light gray color shows the uncertainty on each data
            point. Panels (b--f) show the time (UT) in 20-second intervals.}
            \label{fig:cont_typeI_ds_sizes}
        \end{figure}
        
        Figure \ref{fig:typeI_sizes} shows the cleaned and
        de-convolved FWHM major sizes of the continuum across the almost two-hour observing period across frequency channels with an
        $\sim$1~MHz bandwidth and time averaged every 15.7~s. The data are represented as a 2D
        histogram, and as individual 1D histograms in Figure
        \ref{fig:sizes_1Dhist}. The average size is
        8\arcmin at 31.3~MHz, reducing to 4.3\arcmin at 38.4~MHz. The
        sizes are less than half the size of typical type III bursts
        at the same frequency using collated data presented in
        \cite{2019ApJ...884..122K}. At the same time, some higher
        frequency type I size measurements above 200~MHz appear to
        share a consistent trend if extrapolating the LOFAR
        measurements, whilst many other type I sizes 
        span the range up to that measured for type III bursts.
    
    \subsection{Evolution on Minute Timescales}\label{sec:evolution_minutes}

        Figure \ref{fig:source_pos_time} panels (b--e) show the
        centroid motion over ten minute intervals. In panel (b), the
        centroid motion includes a section that is close to linear,
        moving towards the east. Over this interval, the source shifts
        by $\sim150\arcsec$ over 160~s, corresponding to a projected
        sky-plane velocity of $v\simeq1\arcsec$~s$^{-1}\simeq700$~km
        s$^{-1}$. Over a six minute period, the dynamic spectrum presents several individual type I bursts (e.g. Figure
        \ref{fig:cont_typeI_ds_sizes}). Focusing along a single
        frequency channel at 36.73~MHz (along the dashed horizontal
        line in panel (b)), panels (c--f) highlight 11 type I bursts
        by the vertical dotted lines. From panel (c), the bursts range from a few tens to $\sim100$~sfu brighter than the continuum, which itself is slowly decreasing in brightness from $\sim300$~sfu to $<200$~sfu. Panel (d) shows that the
        continuum source is moving along the horizontal axis of the
        sky-plane beginning near $350\arcsec$ and ending closer to
        $200\arcsec$, whilst the vertical position is stable within
        the uncertainty of a few tens of arcseconds. Some type I
        bursts show a small change in horizontal position (e.g. burst
        number 3), whilst the majority fluctuate within the noise and
        so appear co-spatial with the continuum. In both panels (e)
        and (f), there is generally no change in size between the
        continuum and a type I burst. At the times of bursts 8 and 9,
        there are increases in the major axis size, which could be
        related to the broadband emission that the two type I bursts
        are overlaid onto. Notably, the position of this emission
        makes a minor jump in the opposite direction to most type I
        bursts and includes a small increase in the vertical position.
        The broadband emission could resemble a type III burst along
        an open field line to the north of the active region.
 
        \begin{figure}[htb!]
            \centering
            \includegraphics[width=0.32\textwidth]{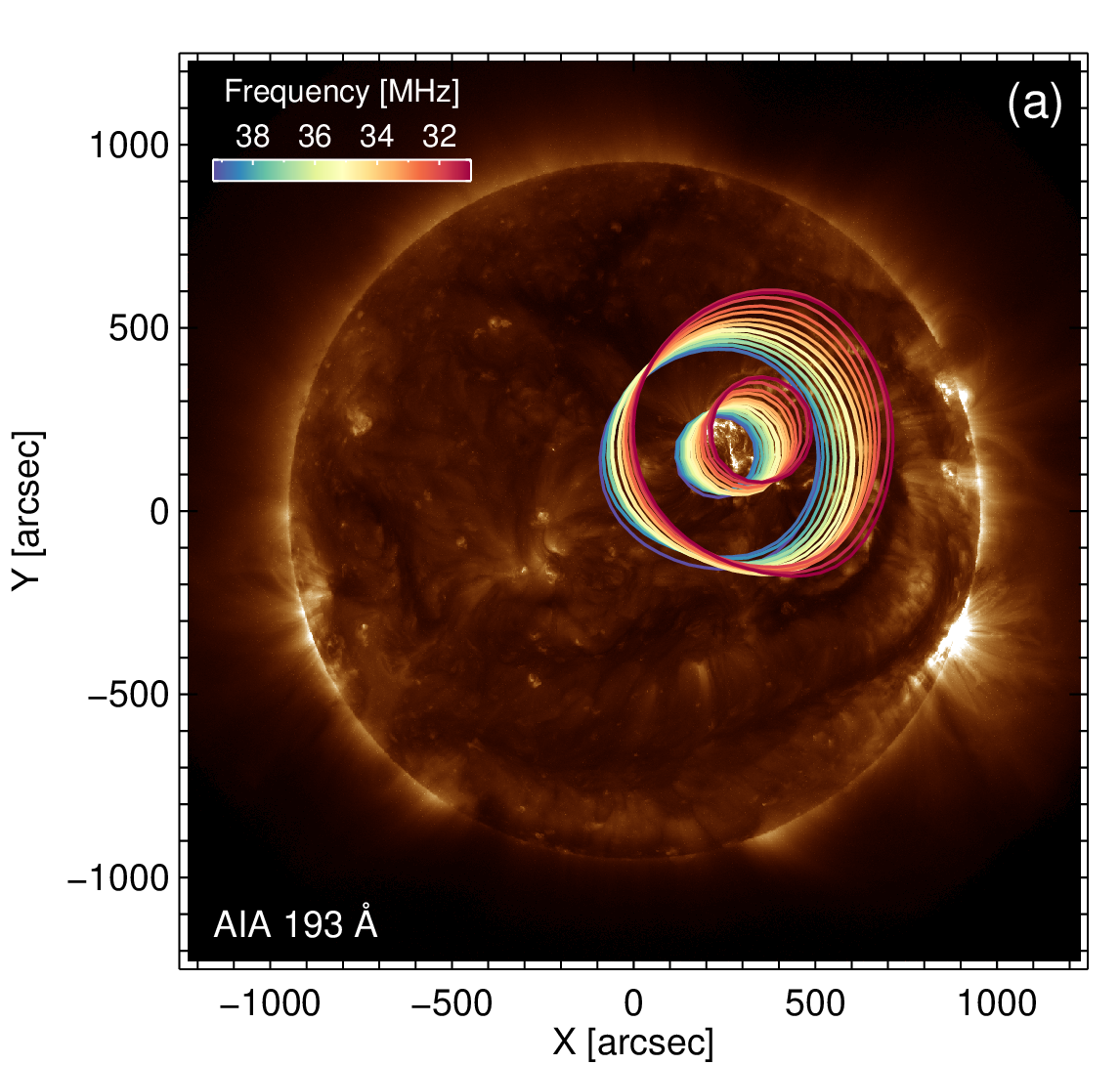}
            \includegraphics[width=0.32\textwidth]{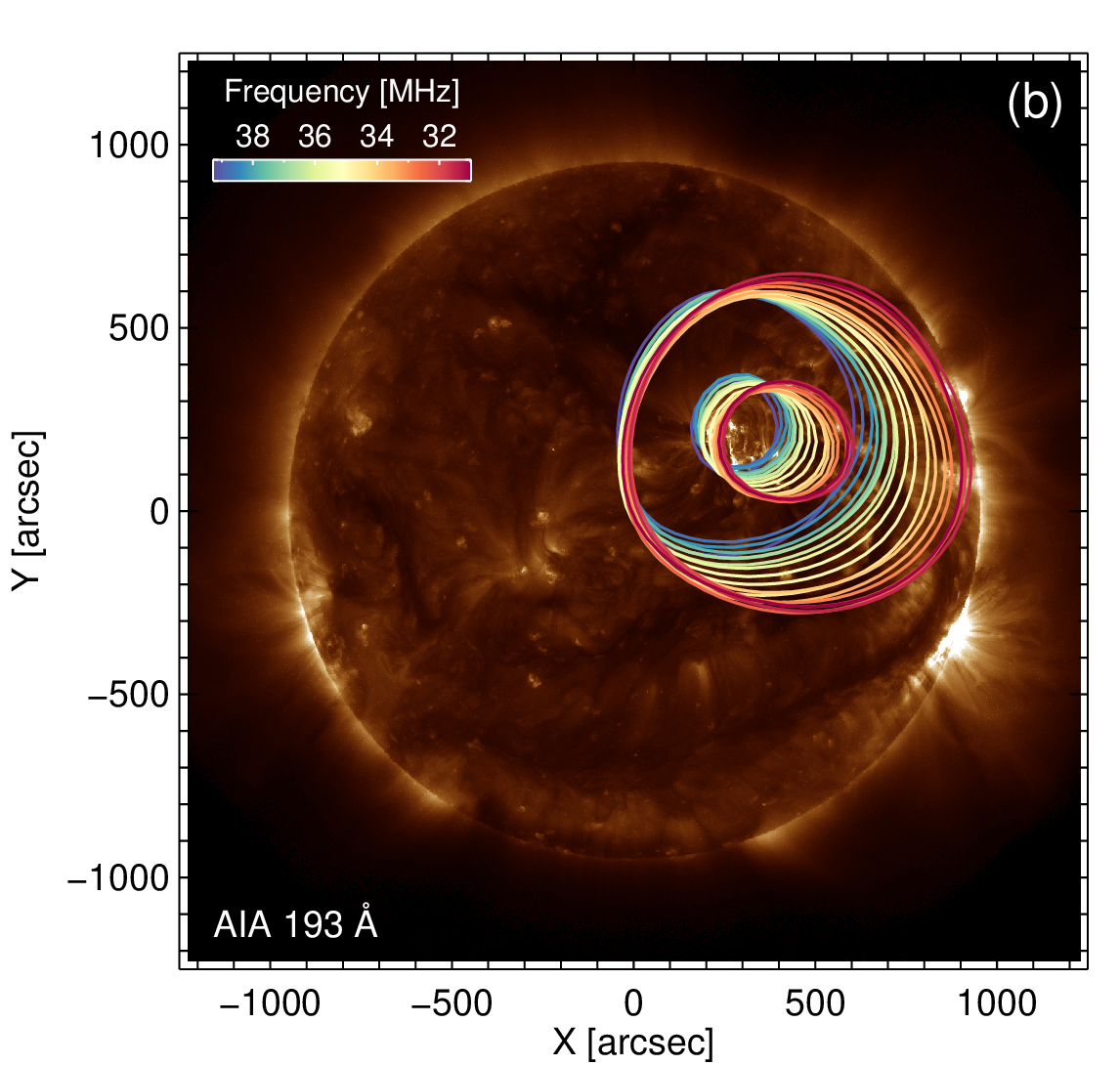}
            \includegraphics[width=0.32\textwidth]{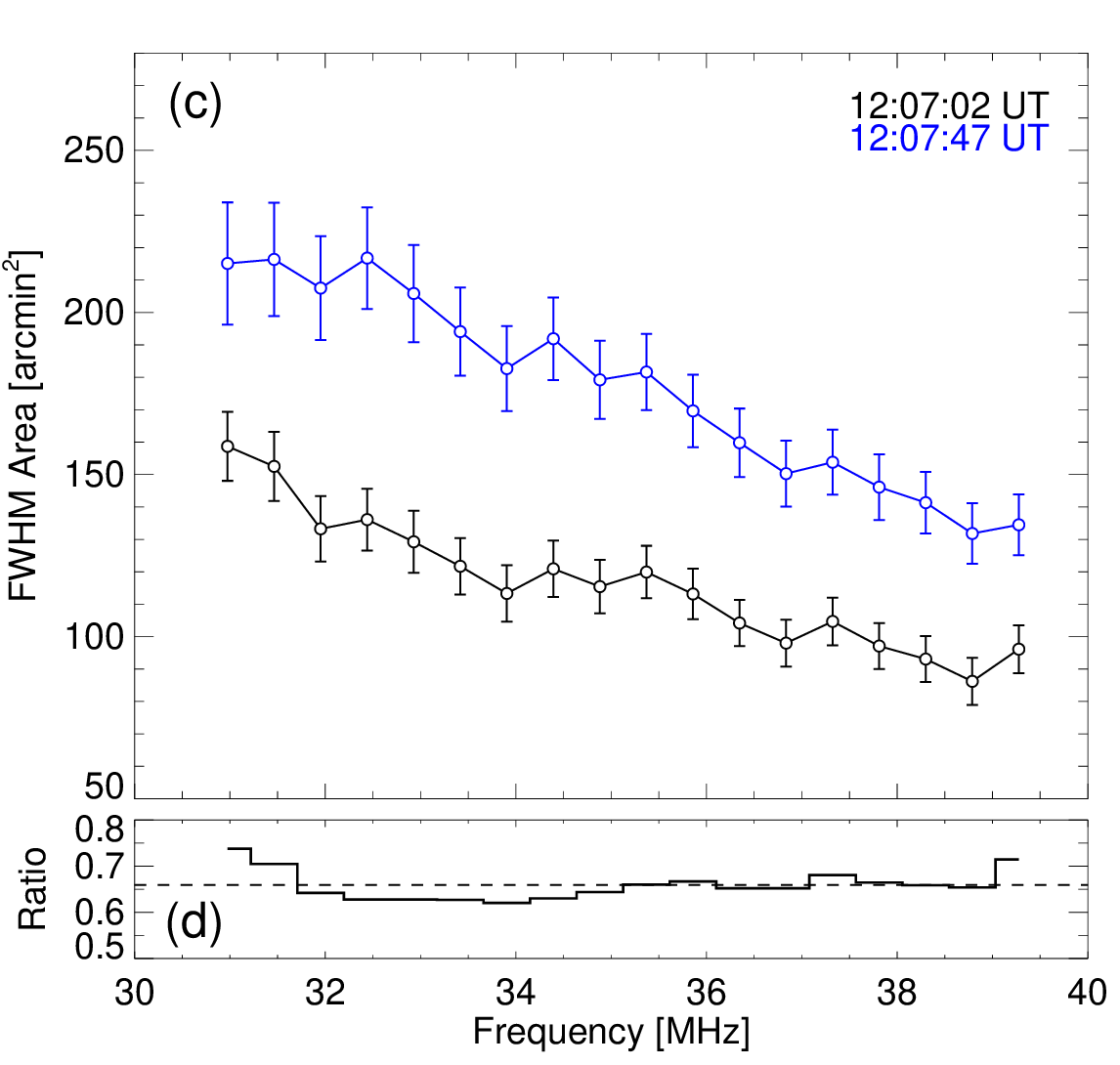}
            \caption{Contours of the cleaned LOFAR images at (a)
            12:07:02 and (b) 12:07:47 UT (white dashed lines in Figure
            \ref{fig:ds_2hr_goes_aia}b and \ref{fig:ds_zoom_flare}) at the 50\% and 90\% intensity
            level, overlaid on an SDO/AIA 193\,\AA\, image at 12:05:06
            UT. The colors map to different frequencies between
            31--39.3~MHz. (c) FWHM area (beam-convolved) against
            frequency from 2D Gaussian fits to each radio map for each
            time. (d) Ratio of the two areas at each frequency, with
            the black dashed line showing the average ratio across all
            frequencies.}
            \label{fig:clean_im_freq}
        \end{figure}
        
        Another notable feature in the dynamic spectrum is the
        fluctuation in frequency of a constant flux density level over
        time, particularly evident before the second flaring episode.
        As shown in Figures \ref{fig:ds_2hr_goes_aia}(b) and
        \ref{fig:ds_zoom_flare}, after 12:00 UT, a flux density level
        of $\sim150$~sfu shifts from 33--34~MHz to near 42 MHz over
        three minutes, before dropping below 30 MHz and undergoing
        repeated oscillations during the following seven minutes. This
        change in the lower frequency extent of the continuum occurs
        roughly ten minutes before the second RHESSI-detected flare.
        At this stage, the GOES soft X-ray flux shows a small rise and
        fall before its main increase at flare onset, at a time when
        the active region exhibits loop brightening (see section
        \ref{sec:event}). The timing also coincides with the
        appearance of the new continuum source (section
        \ref{sec:evolution_2hr}). We additionally investigate the
        source sizes during this period at two times separated by 45~s
        (Figure \ref{fig:clean_im_freq} (a) 12:07:02 UT and (b)
        12:07:47 UT) prior to the flaring episode, between 30--40~MHz.
        These times and frequencies are highlighted in Figure
        \ref{fig:ds_2hr_goes_aia}b and \ref{fig:ds_zoom_flare} by the
        white vertical dashed lines. The cleaned radio images in
        Figure \ref{fig:clean_im_freq} are shown with contours at 50\%
        and 90\% of the peak intensity. Panel (c) shows the radio lobe
        area as a function of frequency, measured at the FWHM level
        using 2D elliptical fits (equation \ref{eq:2dgauss}). As
        expected, the apparent size decreases with increasing
        frequency. However, the earlier time shows consistently
        smaller sizes, with an average reduction by a factor of 0.65
        across the frequency range. Appreciating that these apparent
        sizes are significantly larger than any intrinsic emission
        region due to propagation effects, these variations are
        dominated by changes to the scattering process such as an
        increase in the level of turbulence leading up the flaring
        episode, which would increase the scattering rate per unit
        distance/time, or restructuring of the magnetic field
        topology, affecting the ease of escape of radio-waves within
        anisotropic turbulence. As shown in section
        \ref{sec:sim_results}, field topology can drastically change
        the size of apparent radio sources to a given observer.
     
    \subsection{Second and Sub-second Variability}
    
        \begin{figure}[t!]
            \centering\includegraphics[width=0.45\textwidth]{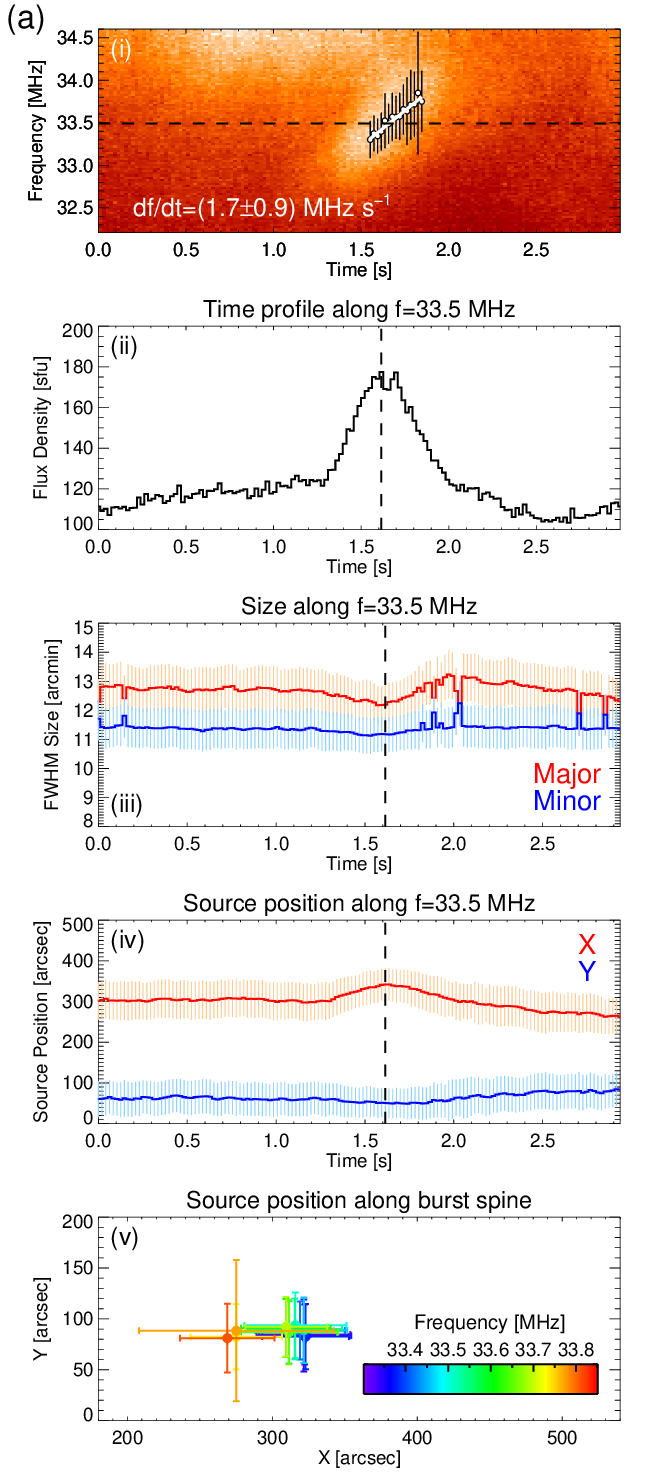}
            \centering\includegraphics[width=0.45\textwidth]{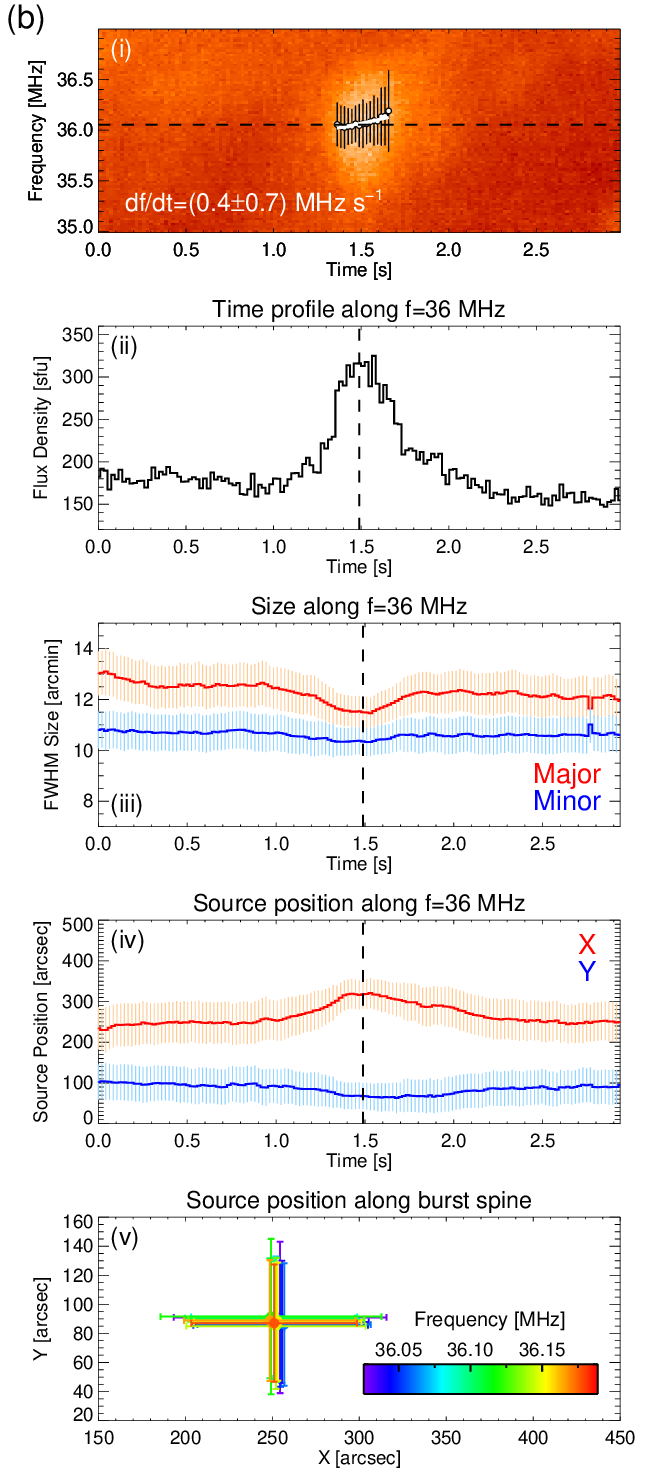}
            \caption{Examples of type I burst drift rates at (a) 10:49:52 UT and (b) 10:50:37 UT. (i) Dynamic spectra. The blue line represents a linear fit to the peak intensities (black) measured every 122~kHz. The dashed line highlights the frequency corresponding to the panels (ii-iv). (ii) Time profiles. (iii) FWHM major and minor sizes over time at a fixed frequency. (iv) Source positions over time at a fixed frequency. (v) Source positions marked by the fitted centroids along the burst spine (i.e. along the linear fit shown in panel (i).}
            \label{fig:typeI_dfdt_size}
        \end{figure}
        
        On sub-second scales, the individual fine structures become
        the most notable features. Figure \ref{fig:typeI_dfdt_size}
        shows two individual type I bursts. From linear fits to the
        time and frequency positions of the peak intensities, their
        drift rates are $(1.7\pm0.9)$~MHz~s$^{-1}$ and
        $(0.4\pm0.7)$~MHz~s$^{-1}$. Using the density model of
        equation \ref{eq:nparker} with the relation
        $\mathrm{d}f/\mathrm{d}t = (\mathrm{d}f/\mathrm{d}r)
        (\mathrm{d}r/\mathrm{d}t)$, this corresponds to exciter
        velocities $v=\mathrm{d}r/\mathrm{d}t$ of approximately
        $|v|=(0.058\pm0.03)c$ and $|v|=(0.011^{+0.021}_{-0.011})c$,
        respectively. Figure \ref{fig:typeI_dfdt} shows the drift
        rates of 12 individual type I bursts from the three minute
        time interval shown in Figure \ref{fig:cont_typeI_ds_sizes},
        with absolute values spread between 0.2--3.4~MHz s$^{-1}$. The
        majority presented positive drift rates, with one third having
        a negative drift rate. Comparatively, the 11 S-bursts
        presented in Figure \ref{fig:sbursts_spikes_ds} have drift
        rates between $-2$ and $-5$~MHz~s$^{-1}$, consistent with
        those measured by \cite{2015A&A...580A..65M}, and similar to
        the fastest drifting type I bursts; however, all of the
        S-bursts present negative drift rates. The 12 spikes analyzed
        in Figure \ref{fig:sbursts_spikes_ds} have absolute drift
        rates between 0.02 and 0.2~MHz~s$^{-1}$, slower than the
        lowest type I burst drift rates, and comparative to the spikes
        measured by \cite{2023ApJ...946...33C} across the same
        frequencies.
        
        \begin{figure}[htb!]
            \centering
            \includegraphics[width=1\textwidth]{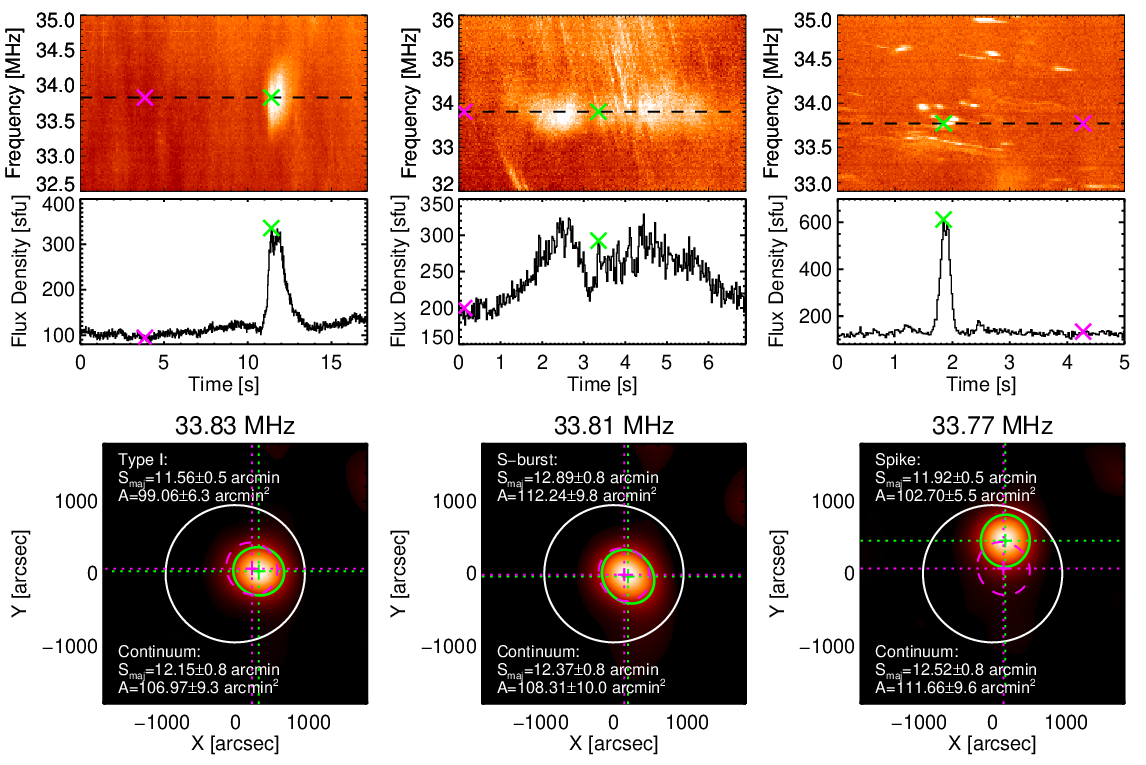}
            \caption{Dynamic spectra, time profiles, and cleaned images of a type I burst (left column), an S-burst (center column), and a spike (right column). From left to right, the dynamic spectra and time profiles begin from 10:50:56.1 UT, 11:22:39.9 UT, and 11:00:00.7 UT, respectively.} Each image is fitted with a 2D Gaussian to determine the major size and area (beam convolved). The green ellipses and crosses mark FWHM fitted contour and the centroid locations of the fine structures, whilst the dashed magenta ellipses and crosses show the FWHM fitted contour and centroid of the continuum. For each case, the colored crosses in the dynamic spectra and time profile denote the time of each image. The dotted lines in the images extend the position of the centroids for visibility.
            \label{fig:fine_struc_cont_size_pos}
        \end{figure}
        
        Figure \ref{fig:typeI_dfdt_size} also presents the major and
        minor sizes and $x$ and $y$ sky-plane positions for the two
        type I bursts. The cleaned size is shown as a function of time
        along the dashed horizontal lines in the dynamic spectra. As
        in Figure \ref{fig:cont_typeI_ds_sizes}, there is little
        change in the size between continuum and type I burst
        emission. At the peak of each burst, there is a reduction in
        the major size, followed by an increase throughout the decay,
        returning back to the continuum level. However, this change is
        within the uncertainty of the measurements. The sources are
        off-center of the solar disk, near 325\arcsec and 250\arcsec
        along the \textit{x}-direction, respectively, and
        80--100\arcsec along the \textit{y}-direction, close to the
        active region shown in Figure \ref{fig:ds_2hr_goes_aia}. Both
        the continuum and bursts are effectively co-spatial within the
        uncertainty of the measurements. Moreover, the extended
        apparent sources of each component overlap substantially
        beyond their centroids, making spatial separation of the two
        difficult. We also present individual type I burst motion
        along the spine of the fine structures. For the example in Figure \ref{fig:typeI_dfdt_size}
        column (a), the source position shifts horizontally by tens of
        arcseconds across the sky-plane, whereas the burst in column
        (b) shows no motion within $\sim40\arcsec$.
    
        In Figure \ref{fig:fine_struc_cont_size_pos}, we compare the
        fine structure images with the fitted size and position of
        continuum within a few seconds of the burst. All fine
        structure types (a type I burst, an S-burst, and a spike)
        present identical sizes and areas within their respective
        uncertainties. Moreover, in each case, the continuum size and
        areas are also identical. As noted with respect to Figure~\ref{fig:fine_struc_cont_size_pos}, the type I burst position is marginally to the west of the continuum, 
        however, as mentioned above it is
        difficult to disentangled the burst from the underlying continuum. 
        Similarly, the S-bursts and continuum are co-spatial. 
        On the other hand, the spike is located to the north of the nearly continuum, 
        with centroids separated by almost $500\arcsec$, suggesting that the spike sources arise
        from a different magnetic field system to the continuum.
        
\section{Radio-wave Propagation Simulations}\label{sec:sim_results}

    \subsection{Simulation Description}

        \begin{figure}[t!]
            \centering
            \includegraphics[width=0.5\textwidth]{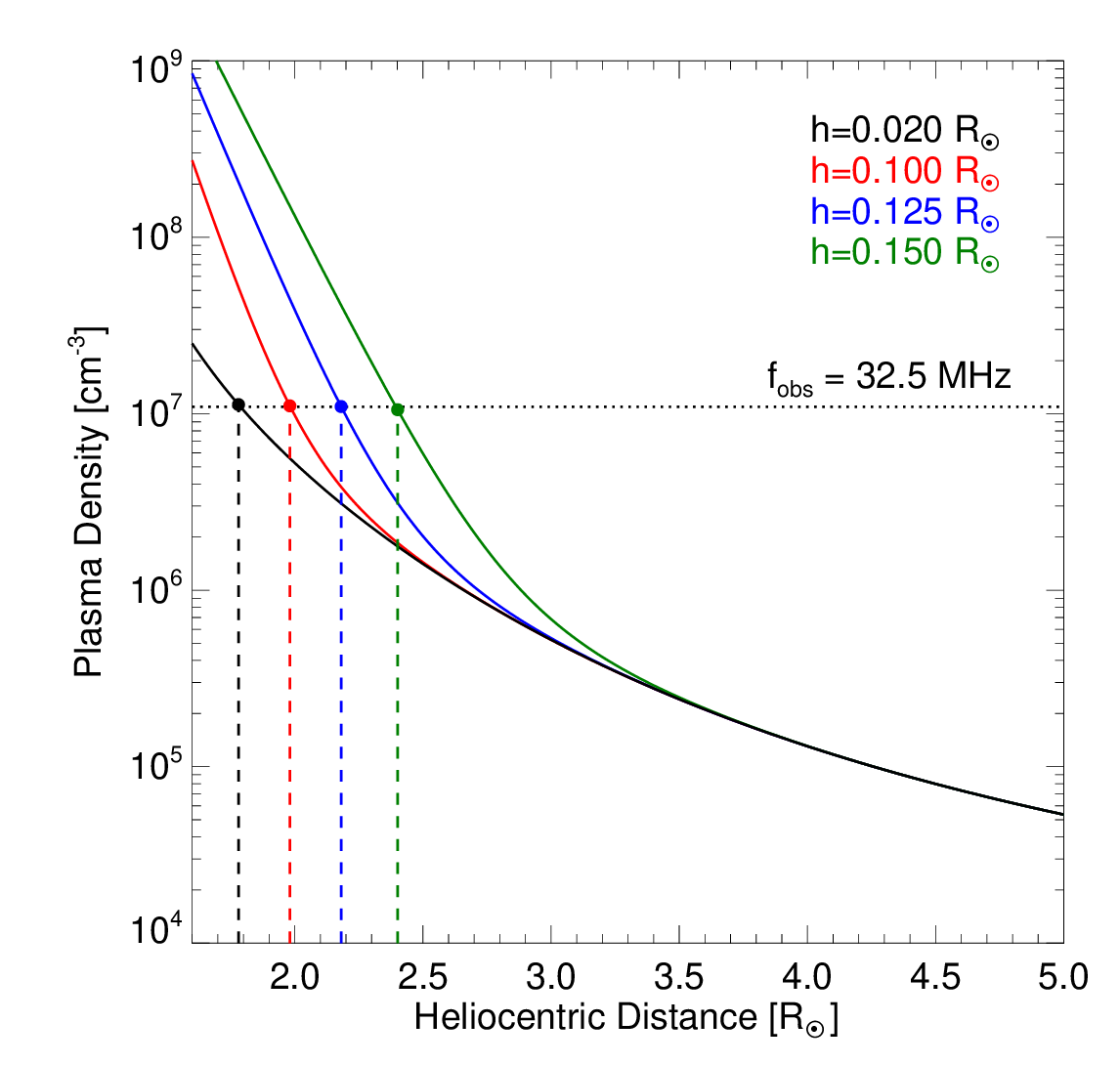}
            \caption{Plasma density profiles described by equation
            \ref{eq:n} for various values of $h$.}
            \label{fig:density_profiles}
        \end{figure}
        
        To probe the required turbulence and field conditions
        necessary to recreate the scattered radio sources, we employ
        the ray-tracing code developed by \cite{2019ApJ...884..122K}
        with a scattering rate that is proportional to a nominal
        heliospheric turbulence profile described in
        \cite{2023ApJ...956..112K} given by
        \begin{equation}\label{eq:qeps2}
            \overline{q\epsilon^2}(r) = \frac{2\times10^3\alpha}{\mathrm{R_\odot}}\left(1-\frac{r}{R_\odot}\right)^{-2.7}\left(\frac{r}{R_\odot}\right)^{-0.7},
        \end{equation}
        where $\epsilon=\langle\delta
        n^2\rangle/n^2=\int{S(\mathbf{q})\,\mathrm{d}^3 q/(2\pi)^3}$
        is the normalized density fluctuation variance, $\mathbf{q}$
        is the wavevector of the density fluctuations, and
        $\alpha=q_\parallel/q_\perp$ describes the anisotropy of the
        density fluctuation spectrum $S(\mathbf{q})$ that is axially
        aligned to the magnetic field. We consider anisotropy factors of 0.1 and 0.25, as well as scaling the
        turbulence profile by a factor of 1 and 1/2. We also consider
        magnetic fields in both open and closed configurations. The
        open field structure is described by a Parker spiral with an
        Alfv\'{e}n radius at $15$~R$_\odot$ given by
        $\mathbf{B}(r)=B_r(r)\hat{r}+B_\theta(r)\hat{\theta}$ with
        components
        \begin{equation}\label{eq:bfield}
            \begin{split}
                B_r(r) &= B_0\left(\frac{215~\mathrm{R_\odot}}{r}\right)^2\\
                B_\theta(r) &= -B_0\left(\frac{215~\mathrm{R_\odot}}{r}\right)\left(\frac{\Omega\left[r - 15\,\mathrm{R}_\odot\right]}{v_\mathrm{sw}}\right)\sin{\phi},
            \end{split}
        \end{equation}
        where $\phi=\pi/2$, $B_0\simeq5\times10^{-5}$~G gives the
        magnetic field strength at 1~au, the solar rotation rate is
        $\Omega=2.7\times10^{-6}$~rad s$^{-1}$ and we use a typical
        slow solar wind speed at 1~au of $v_\mathrm{sw}=420$~km
        s$^{-1}$. The closed field structure is described by a dipole
        as used in \cite{2025ApJ...978...73C}, where
        \begin{equation}\label{eq:dipole_eq}
            \mathbf{B}(r^\prime, \phi^\prime)=\frac{B_0}{(r^\prime)^3}\left(2\cos{(\phi^\prime)}{\hat{r}^\prime} + \sin{(\phi^\prime)}{\hat{\phi}^\prime}\right),
        \end{equation}
        that approximates a coronal loop far from the photosphere. The
        dipole center is placed at a location given by
        $\mathbf{d}=(r_\mathrm{d}, \theta_\mathrm{d},
        \phi_\mathrm{d})$ as described in \cite{2025ApJ...978...73C}
        (see their Figure 1).
        
        We inject photons as a delta function (point source) with
        initially isotropic wavevector $k$ distribution at a height
        $r_0$ with a frequency maintained throughout the simulation
        that is defined by the ratio $f_\mathrm{obs}/f_\mathrm{pe}$,
        where $f_\mathrm{pe}$ is the plasma frequency determined by
        the nominal plasma density model
        \begin{equation}\label{eq:n}
            n(r) = n_p(r) + n_l(r),
        \end{equation}
        where
        \begin{equation}\label{eq:nparker}
            n_p(r) = 4.8\times10^9\left(\frac{r}{\mathrm{R}_\odot}\right)^{-14} + 3\times10^8\left(\frac{r}{\mathrm{R}_\odot}\right)^{-6} + 1.4\times10^6\left(\frac{r}{\mathrm{R}_\odot}\right)^{-2.3}
        \end{equation}
        is an analytical form of that derived by
        \cite{1960ApJ...132..821P} normalized to 1~au with constants
        by \cite{1999A&A...348..614M}, as used in
        \cite{2019ApJ...884..122K}. The term $n_l(r)$ allows for
        increased densities of coronal loops of different heights
        where
        \begin{equation}\label{eq:nc}
            n_l(r) = 10^{11}\exp\left(-\frac{r-1}{h}\right),
        \end{equation}
        and $h=[0.02, 0.1, 0.125, 0.15]$~R$_\odot$ defines the height
        at which the loop density enhancement vanishes to background
        levels. We consider fundamental plasma emission where
        $f_\mathrm{obs}=1.1f_\mathrm{pe}$. To maintain the same
        frequency for each simulation despite differing density
        profiles, the sources are placed at $r_0=[1.79, 1.98, 2.18,
        2.39]$~R$_\odot$, as shown in Figure
        \ref{fig:density_profiles}.
        
        The photons elastically scatter until they reach a radius of
        1~au. The collected photons are back-projected to form an
        $(x,y)$ photon position map, from which a 2D histogram forms
        the scatter image, where each photon $i$ is weighted by
        $w_i=e^{-\tau}$. Here,
        $\tau=\int{\gamma(\mathbf{r}(t))}\,\mathrm{d}t$,
        $\gamma=\left(f_\mathrm{pe}^2/f^2\right)\gamma_c$ is the
        free-free absorption rate, and $\gamma_c$ is as defined in
        \cite{2019ApJ...884..122K}. The scatter images are then fit
        with a 2D Gaussian given by equation \ref{eq:2dgauss} to
        determine their various characteristics.

    \subsection{Simulation Results}

        \begin{figure}[htb!]
            \centering
            \includegraphics[width=1\textwidth]{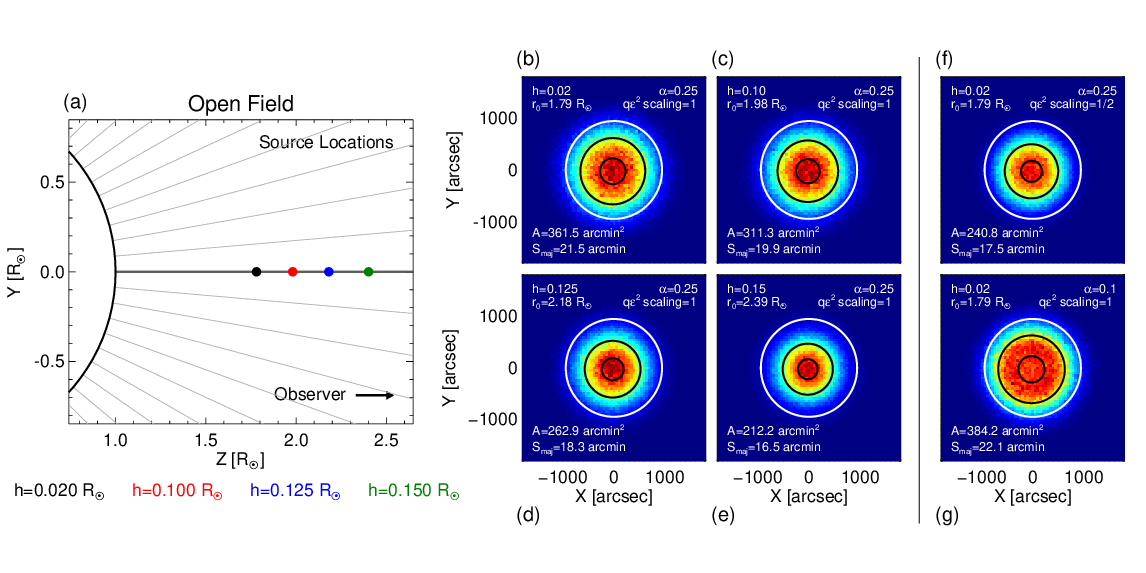}
            \caption{Scatter images for sources placed within an open
            magnetic field structure. (a) Source locations at 32.5~MHz
            (colored circles) for the different density profiles shown
            in Figure \ref{fig:density_profiles}. The partial black
            circle represents the solar disk, and the observer lies
            along the \textit{z}-axis. The black radial line denotes
            the field line at the source latitude. The thin grey lines
            show the radial field at other latitudes. (b)-(e)
            Simulated radio images in the $(x,y)$ plane for each
            source shown in panel (a). The sizes and area are given in
            each panel by $S_\mathrm{maj}$ and $A$, derived from the
            2D fitted Gaussian's, shown by the black circles at the
            50\% and 90\% levels. The white circle represents the
            solar disk. (f-g) Simulated radio images for $h=0.02$ with
            halved scattering rate (f) and stronger anisotropy (g),
            to be compared with panel (b).}
            \label{fig:sim_images_open}
        \end{figure}
        
        \begin{figure}[htb!]
            \centering
            \includegraphics[width=1\textwidth]{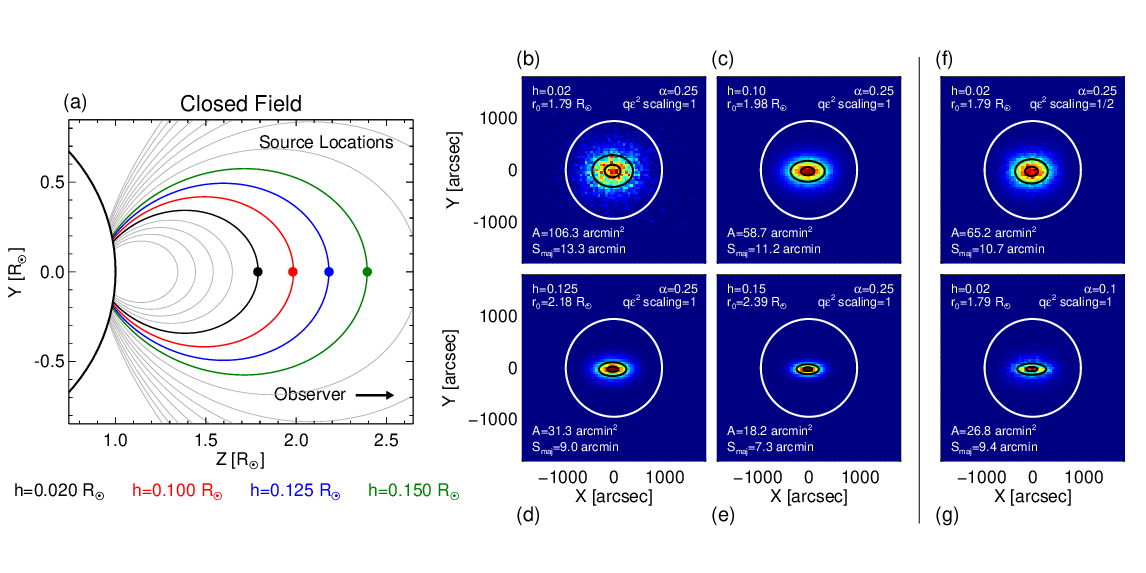}
            \caption{As in Figure \ref{fig:sim_images_open} but for
            sources placed in closed magnetic structures formed by a
            dipole.}
            \label{fig:sim_images_dipole}
        \end{figure}

    In Figures \ref{fig:sim_images_open} and \ref{fig:sim_images_dipole},
    each image shares the same observed frequency of $32.5$~MHz and the
    photon emission sources are placed at the solar disk center. Figure
    \ref{fig:sim_images_open} presents sources placed along an open field
    configuration, whilst the sources for Figure \ref{fig:sim_images_dipole} 
    are located within a closed field
    configuration. For each case, the density profiles follow equations
    (\ref{eq:n})-(\ref{eq:nc}) for the various values of $h$, as displayed
    in Figure \ref{fig:density_profiles}.
    
    For both field configurations, the source sizes decrease for density
    profiles that have a higher density and steeper gradient at a given
    source height. In the open configuration, the sky-plane sources are
    spherical, whilst the closed configuration leads to elliptical
    sources. This is expected since anisotropic scattering leads to broadening preferentially perpendicular to the magnetic field---for
    the closed field structure we considered, the major axis lies along
    the \textit{x}-axis, yet this orientation will be obscured when
    convolved with the instrument beam (see Figure 4 of
    \cite{2025ApJ...978...73C}, for example). The FWHM major sizes in the
    closed field system reduce from $21.5\arcmin$ to $16.5\arcmin$,
    remaining double the size of the measured noise storm sizes as shown
    in Figure \ref{fig:typeI_sizes} and similar to the type III burst
    source sizes. Comparatively, the closed field configuration has
    elliptical sources has approximately half the source sizes as those
    along an open field, reducing from $13.3\arcmin$ to $7.3\arcmin$ for
    the different density profiles. The smallest sizes (relating to
    density profiles using $h=0.125$ and $h=0.15$) show agreement with the
    noise storm sizes measured by LOFAR in Figure \ref{fig:typeI_sizes}.
    We further show the extent of the source size reduction for varying
    both the scattering rate and anisotropy. In the open field case
    (Figure \ref{fig:sim_images_open}), panel (f) presents the scattered
    source for a density profile with $h=0.02$ and half the scattering
    rate. Compared with panel (b), the source size is reduced by 19\%. In
    panel (g), the same scenario for nominal scattering rate and stronger
    anisotropy of $\alpha=0.1$ has eligible effect on the source size, in
    agreement with results in \citep{2023ApJ...956..112K}. The minor
    increase in size by a few percent is likely a focusing effect along
    the field lines directed towards the observers line of sight. For the
    closed field case (Figure \ref{fig:sim_images_dipole}), the same
    panels show that reducing the scattering rate has the same effect as
    in the open field case, however, the stronger anisotropy has a
    marked influence on the source size, reducing it by almost 30\%.
    
    As described in section \ref{sec:lofar_results}, the noise storm
    appears to exist within the large coronal loop system. In such a
    scenario, the density fluctuation anisotropy of the confined plasma
    causes a differential scattering rate with respect to the magnetic
    field. In the open configuration, radio-waves readily escape outwards
    from the Sun and are focused towards the observer. In the closed
    field, the radio-waves are preferentially scattered away from the
    observer in both directions of the magnetic loop, and weaker level
    radio emission escapes the system and reaches the observer.
    Consequently, the source sizes are smaller for radio sources embedded
    in closed magnetic fields compared to open field systems, dependent on
    the position of the observer.
    
    The findings from the simulations highlight why the source sizes for
    the different types of emission analyzed in this work are, within
    their uncertainties, very similar. The observed sizes are dominated by
    radio-wave propagation effects, which itself is coupled to both the
    plasma environment and the magnetic field structure. Open field
    structures directed towards the observer lead to larger sources due to
    the ease of radiation escape to the observer (e.g. type III bursts),
    whilst sources in closed fields lead to smaller sources from weaker
    levels of radiation escape towards the observer. The fine structures
    in this event comprising type I bursts, S-bursts, and spikes are all
    emitted from a complex magnetic environment of open and closed field
    lines, complicating radiation escape with respect to anisotropic
    scattering, and leading to small source sizes comparable to the noise
    storm trapped with the coronal loop. The spread in source sizes of the
    noise storm and type I bursts shown in Figure \ref{fig:typeI_sizes}
    can be replicated in this model by appreciating the different plasma
    conditions and field configuration per event, or during a single
    event.

\section{Discussion}\label{sec:discussion}

    We analyzed LOFAR tied-array imaging of a near disk-center noise
    storm with embedded fine structures (type I bursts, S-bursts, and
    spikes) between 30--40~MHz, and compared the observations with
    anisotropic radio-wave scattering simulations. The observations
    show a compact apparent source that drifts across the solar disk
    over a two-hour interval, consistent with emission confined within
    a complex system of closed magnetic field lines.
    
    Individual type I bursts exhibit frequency drift rates between
    0.2--3.4~MHz~s$^{-1}$. Bursts with very low drift rates are
    ambiguous: depending on the method, they can be interpreted as
    either near-zero drift or effectively infinite drift across a
    finite bandwidth. In our approach, which tracks peak emission in
    frequency over time, such cases yield small drift rates.
    Regardless of interpretation, these bursts imply emission across a
    finite bandwidth of $\sim1$~MHz. Using the density model (equation
    \ref{eq:nparker}), this corresponds to a spatial extent of
    $\sim0.013$~R$_\odot$ ($\sim10^9$~cm), comparable to acceleration
    region sizes inferred for type III bursts
    \citep{2011A&A...529A..66R}, and an order of magnitude larger than
    those associated with radio spikes \citep{Clarkson_2021}.

    If the continuum emission arises from a trapped electron
    population, as proposed by \cite{2017SoPh..292..117L}, then the
    presence of fine structures with distinct temporal and spectral
    properties implies an additional emission process. The drift rates
    of the S-bursts are as high as the fastest type I drifts whilst
    the spikes drift rates are as low as the slowest drifting type I
    bursts. Both these types of fine structures, at decameter
    frequencies, are attributed to fast electron beams and the plasma
    emission process
    \citep[e.g.][]{Clarkson_2021,2014SoPh..289.1701M}. A natural
    interpretation is that the type I bursts are also produced by
    locally accelerated electron beams within the same coronal
    environment. Their close spatial association with the continuum
    suggests that this acceleration occurs within or near the same
    magnetic structure, likely driven by small-scale magnetic
    reconnection.

    Across the observed frequency range, the major axis of the noise
    storm decreases from $\sim8.0\arcmin$ at 31.3~MHz to
    $\sim4.3\arcmin$ at 38.4~MHz, approximately half the typical size
    of type III burst sources at comparable frequencies. At a fixed
    frequency, the apparent sizes of type I bursts, S-bursts, and
    spikes remain consistent with the underlying continuum. On
    sub-second timescales, a small decrease in the type I burst source
    size is observed at the burst peak, followed by a return to the
    continuum level during decay. However, since the burst intensity
    is typically only 50\%--80\% above the continuum, the measured
    source represents a convolution of both components. The apparent
    size variation may therefore be driven by changes in the relative
    brightness weighting in the image-plane rather than a true change
    in the intrinsic emission region.

    The simulations demonstrate that the observed compact source sizes
    are dominated by propagation effects rather than intrinsic source
    properties. Since intrinsic emission regions are orders of
    magnitude smaller than the observed images
    \citep[e.g.][]{2017NatCo...8.1515K, 2019ApJ...884..122K,
    Clarkson_2021}, the apparent size is governed by the plasma
    environment and magnetic field configuration, rather than the
    burst type itself. In particular, compact apparent sources arise
    from a combination of:
    \begin{enumerate}
        \item Closed magnetic field structures, which limit the escape
        paths of radiation toward the observer.
        \item Over-dense plasma regions, which steepen density
        gradients and modify the effective scattering region.
        \item Reduced turbulence levels, which lower the scattering
        rate.
        \item Strong anisotropy in the density fluctuation spectrum
        in non-radial (or closed) field structures, with $\alpha\simeq0.1$.
    \end{enumerate}
    
    Among these factors, the magnetic field configuration has the
    strongest effect, and is strongly linked to the anisotropy of the
    turbulence spectrum. In anisotropic turbulence, where density
    fluctuations are elongated along the magnetic field, stronger
    anisotropy focuses radiation along the field by increasing the mean
    free path parallel to the field direction. As a result, radiation
    preferentially diffuses along magnetic field lines. In open field
    geometries, where the field is approximately radial and aligned with
    the observer's line of sight, this leads to efficient transport of
    radiation toward the observer and a broader angular distribution on
    the sky, producing larger apparent source sizes. Coronal loops, where
    the plasma is confined within closed field structures, have previously
    been suggested to have stronger anisotropy than open field lines (e.g.
    \cite{Clarkson_2021, 2023ApJ...946...33C}). In the latter study,
    reproducing the rapid, non-radial apparent motion of spike sources
    required strong anisotropy between $\alpha=0.1-0.2$. Moreover,
    simulations by \cite{2020ApJ...898...94K}, which required strong
    anisotropy to reproduce the double-peaked structure of drift-pair
    bursts through turbulent echos, show a more pronounced effect at lower
    values of $\alpha$. In their Figure 4, the apparent source area at the
    \textit{peak} flux density is reduced by $\sim25\%$ when $\alpha=0.1$
    compared with $\alpha=0.2-0.3$. For sources embedded near the tops of
    closed loops, the field direction is largely transverse to the
    observer. The strong anisotropy then guides the radiation away from
    the line of sight, and the observer receives emission from a more
    restricted angular range. This reduces not only the received flux, but
    also the angular extent of the detected emission, resulting in more
    compact apparent sources. Comparison of Figure
    \ref{fig:sim_images_open} panels (b) and (g) shows that, in open field
    configurations, image sizes integrated over the \textit{full} time
    profile change little between $\alpha=0.25$ and $\alpha=0.1$. In
    closed field structures (Figure \ref{fig:sim_images_dipole}), however,
    the effect is much stronger. We find a reduction in apparent source
    size of about $4\arcmin$ caused by the increased directivity along
    field lines allowing for bulk radiation escape away from the observer.
    
    \cite{2023ApJ...956..112K} show that reducing the turbulence level by
    a factor of two decreases the apparent source size from approximately
    22$\arcmin$ to 18$\arcmin$ at 30~MHz. Similarly, to explain the
    relatively small sizes of radio spikes confined within a coronal loop,
    \cite{2023ApJ...946...33C} required the turbulence level to be reduced by almost 1/2,
    corresponding to a halving
    of the scattering rate. Consistent with these results, Figures
    \ref{fig:sim_images_open}(f) and Figure \ref{fig:sim_images_dipole}(f)
    shows that halving the scattering rate reduces the apparent source
    size by $\sim20\%$ for both open and closed field sources. We also
    find a reduction in source size when comparing the radio images formed
    within the same magnetic structures but with increased plasma density.
    A higher density produces a steeper density gradient in the source
    region before returning to background levels, placing the emission at
    a given frequency farther from the Sun. Whilst the strongest
    scattering occurs close to the plasma frequency surface,
    \citep{2023ApJ...956..112K} show that rapid scattering occurs within a
    spherical region around the source location. For example, at 30~MHz
    emission, strong scattering occurs in a shell of $\sim1$~R$_{\odot}$
    in radial extent. If the density gradient is steeper, this shell
    becomes narrower, so strong scattering is confined to a smaller
    region, thereby reducing the amount of source broadening. As shown in
    Figures \ref{fig:sim_images_open} and \ref{fig:sim_images_dipole},
    increasing the density of a region by $\sim70\%$ produces only a
    modest reduction in source size, of less than $2\arcmin$. A more
    extreme density enhancement of an order of magnitude reduces the
    apparent source size by approximately 25\% in open field structures,
    and 45\% in closed structures. It is worth noting that the density
    gradient plays an important role in type II bursts.
    \citet{2018ApJ...868...79C} show how the density jump or strong
    density gradient leads to the apparent spatial displacement between
    split-band type II sources. The density jump across the shock produces
    two plasma frequencies both near local plasma frequencies, but because
    the lower-frequency (upstream) emission is closer to unperturbed
    background density the radio-waves are scattered stronger than the
    higher-frequency emission from the downstream, higher density region.
    
    We note that \cite{2015A&A...576A.136M} argue that the smallest
    observed source sizes place upper limits on scatter broadening and may
    indicate that scattering effects have been overestimated. An
    alternative perspective emerges, however, when considering that these
    constraints are largely inferred from the minor axis of elongated
    sources observed near the solar limb. In the presence of anisotropic
    scattering, the apparent source structure reflects the angular
    distribution of radiation at the surface of last scattering, which
    depends on both the magnetic field orientation and the observer's
    viewing angle. For limb sources, the observer samples emission from
    the side of this distribution, where the projected extent can be
    significantly reduced even when scattering remains strong. From this
    viewpoint, the minor axis may not provide a direct measure of the
    intrinsic scatter-broadening, and could instead underestimate the true
    extent of the scattered emission. Consequently, the observed compact
    sizes may still be consistent with substantial scattering under
    anisotropic conditions. Moreover, as shown in Figure
    \ref{fig:sim_images_dipole}, anisotropic scattering for sources in
    coronal loops produces elliptical sources. At decameter frequencies,
    such structure is difficult to observe due to convolution with the
    instrument beam, yet there are examples at decimeter frequencies where
    elongated sources appear between sunspot groups, with the major axis
    perpendicular to possible connecting field lines 
    \cite[see Figure 4 of][for example]{1987ApJ...319..514L}.
    
    These results provide a consistent explanation for the observed
    similarity in source sizes across different fine structure types.
    Despite their distinct spectral and temporal properties, type I
    bursts, S-bursts, and spikes share the same propagation environment,
    and therefore exhibit similar apparent sizes. The spread in observed
    sizes across events (Figure \ref{fig:typeI_sizes}) can be understood
    as arising from variations in magnetic topology, density structure,
    and turbulence conditions, rather than intrinsic differences between
    emission mechanisms.
    
\section{Conclusion}\label{sec:conclusion}
        
    We have presented LOFAR imaging spectroscopy of a solar radio noise
    storm with embedded fine structures at 30--40~MHz, combining high time
    and frequency resolution observations with radio-wave propagation
    simulations. The noise storm forms a compact radio source that drifts
    across the solar disk over timescales of tens of minutes, consistent
    with motion within a closed magnetic field system. 
    The embedded type I bursts exhibit a wide range of frequency drift 
    rates (0.2--3.4~MHz~s$^{-1}$), 
    with S-bursts consistent with the fastest
    drifting type I bursts, and spikes consistent with the slowest
    drifting type I bursts. Their distinct spectral and temporal
    properties suggests the fine structures are generated by plasma
    emission from accelerated electron beams, likely driven by magnetic
    reconnection in the vicinity of the noise storm source. Most fine
    structures are co-spatial with the continuum, and all share compact
    apparent sizes identical within the uncertainties (significantly smaller
    than those of type III bursts), indicating that their
    observed properties are dominated by propagation effects rather than
    intrinsic source sizes. 
    While turbulence levels are expected to influence
    the degree of scatter broadening leading to larger/smaller source sizes, 
    our simulation results show that geometric effects related to the magnetic field configuration in anisotropic turbulence substantially modify the apparent source
    sizes, producing rather compact sources even in the presence 
    of significant scattering.
    
    The simulations accounting for the magnetic field structures 
    show that these compact apparent sizes arise primarily from emission 
    within closed magnetic field structures, where anisotropic scattering directs radiation away
    from the observer's line of sight.
    Additional factors such as plasma density gradients further influence 
    the observed sizes, but remain secondary to magnetic geometry in anisotropic plasma.
    These results demonstrate that the apparent size of
    solar radio sources at decametric wavelengths is governed mainly by
    the large-scale coronal environment rather than the intrinsic emission
    process, providing a unified explanation for the similar sizes of
    diverse fine structures within noise storms and highlighting the
    importance of magnetic topology and plasma conditions in shaping solar
    radio observations.
    
\begin{acknowledgments}
    The work is supported by UKRI/STFC grant ST/Y001834/1. 
    EPK is partially supported by the Leverhulme Trust 
    (Research Fellowship RF-2025-357). The authors
    acknowledge the support by the international team grant
    (\href{http://www.issibern.ch/teams/lofar/}{http://www.issibern.ch/teams/lofar/})
    from ISSI Bern, Switzerland. This paper is based (in part) on data
    obtained from facilities of the International LOFAR Telescope
    (ILT) under project codes LC3\_012 and LC4\_016. LOFAR
    \citep{2013A&A...556A...2V} is the Low-Frequency Array designed
    and constructed by ASTRON. It has observing, data processing, and
    data storage facilities in several countries, that are owned by
    various parties (each with their own funding sources), and that
    are collectively operated by the ILT Foundation under a joint
    scientific policy. The ILT resources have benefited from the
    following recent major funding sources: CNRS-INSU, Observatoire de Paris and Universit\'{e} d'Orl\'{e}ans, France; BMBF, MIWF-NRW,
    MPG, Germany; Science Foundation Ireland (SFI), Department of
    Business, Enterprise and Innovation (DBEI), Ireland; NWO, The
    Netherlands; The Science and Technology Facilities Council, UK;
    Ministry of Science and Higher Education, Poland.
\end{acknowledgments}

\appendix
\counterwithin{figure}{section}
\renewcommand\thefigure{\thesection\arabic{figure}}

    \section{Frequency Drift Rates of Type I Bursts, S-bursts and Spikes}

\begin{figure}[htb!]
        \centering
        \includegraphics[width=1\textwidth]{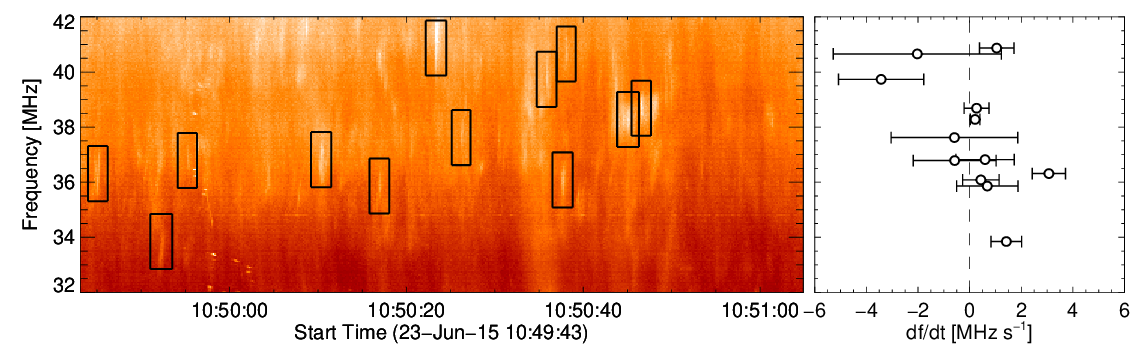}
        \caption{Left panel: Dynamic spectra showing numerous type I bursts overlaid on continuum emission. Time (UT) is shown in 20-second intervals. The black boxes highlight bursts whose frequency drift rate is measured. Right panel: Absolute frequency drift rate as a function of frequency for the bursts fitted in the left hand panel.}
        \label{fig:typeI_dfdt}
    \end{figure}

    Figure \ref{fig:typeI_dfdt} shows a 1.3~minute dynamic spectrum of the noise storm with numerous type I bursts highlighted by the 12 black squares. For each burst, the type I frequency drift rate is measured as detailed in section \ref{sec:data_methods}. The resulting drift rates are shown in the right hand panel with a spread from $-3.4$~MHz~s$^{-1}$ to $+3.1$~MHz~s$^{-1}$.
        
\begin{figure}[htb!]
        \centering
        \includegraphics[width=1\textwidth]{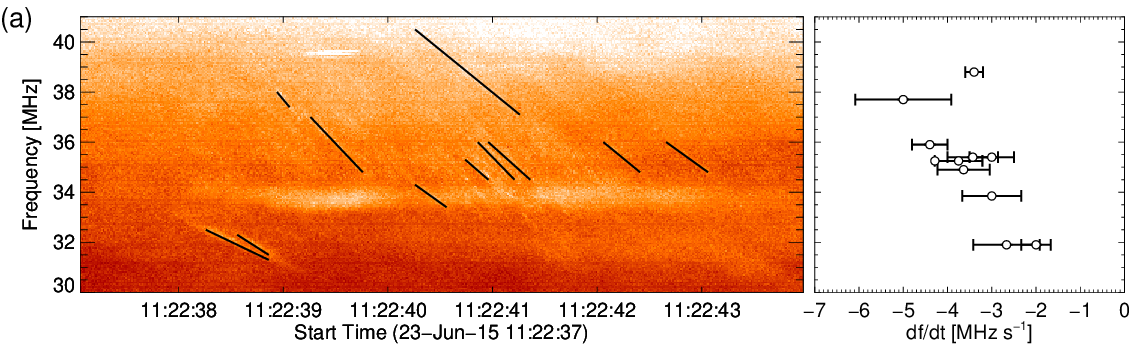}
        \includegraphics[width=1\textwidth]{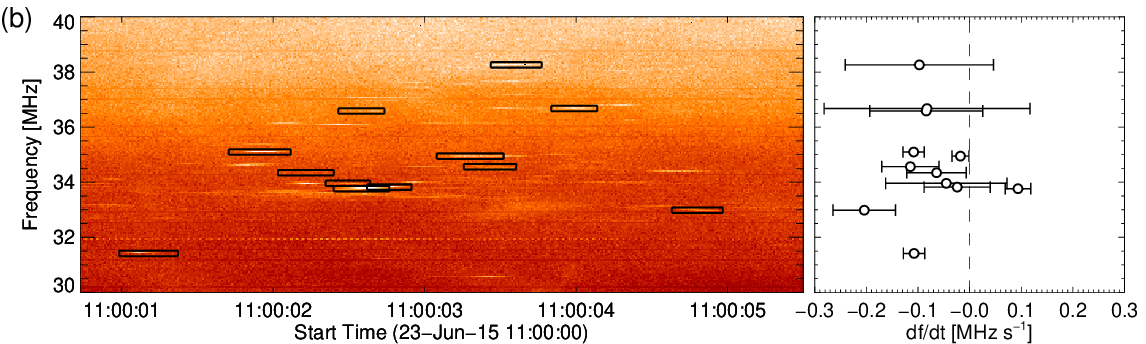}
        \caption{Left column: Dynamic spectra showing (a) S-bursts and (b) spikes. Time (UT) is shown in 1-second intervals. Right column: Frequency drift rates of bursts highlighted in the dynamic spectra.}
        \label{fig:sbursts_spikes_ds}
    \end{figure}

    Figure \ref{fig:sbursts_spikes_ds} shows examples of other fine structures embedded in the noise storm. The 11 S-bursts in
    row (a) highlighted with the black lines have a spread in drift rate between $-5$~MHz~s$^{-1}$ and $-2$~MHz~s$^{-1}$. The 12 spikes in row (b) are bounded by the black boxes and have drift rates between $-0.2$~MHz~s$^{-1}$ and $+0.1$~MHz~s$^{-1}$.
        
\section{Dynamic Spectrum During the Flaring Time Period}

\begin{figure}[htb!]
        \centering
        \includegraphics[width=0.8\textwidth]{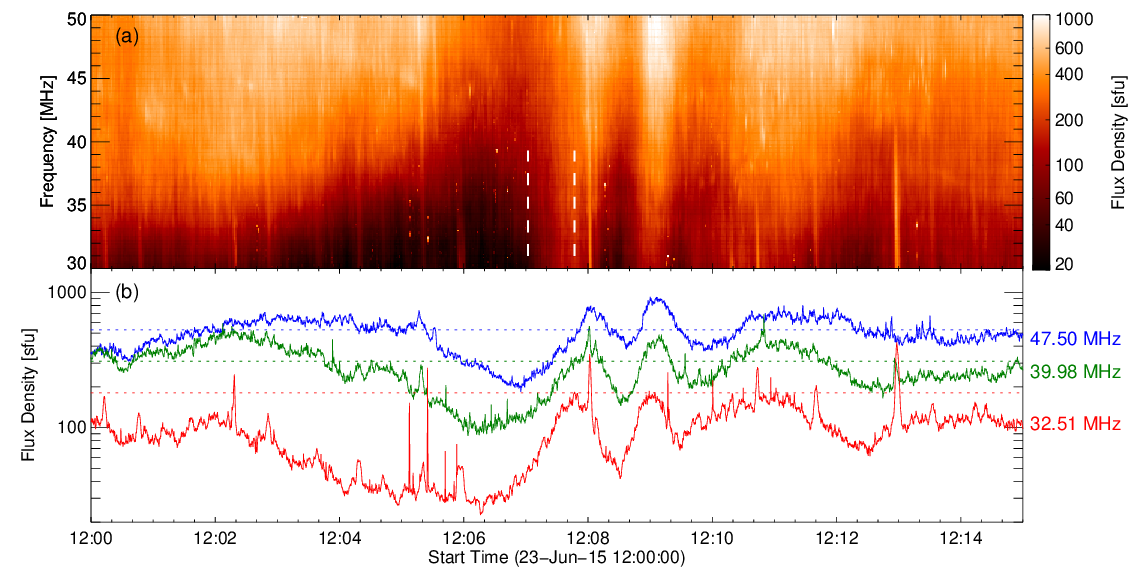}
        \caption{(a) Zoomed in dynamic spectra of the region
        highlighted in Figure \ref{fig:ds_2hr_goes_aia}. The two
        vertical dashed white lines denote the time and
        frequencies of the images in Figure
        \ref{fig:clean_im_freq}. (b) Lightcurves from the spectrum
        in panel (a) at three frequencies. The dotted lines show
        the average flux at these frequencies before 12:00 UT.
        Time (UT) is shown in 2-minute intervals.}
        \label{fig:ds_zoom_flare}
    \end{figure}

    Figure \ref{fig:ds_zoom_flare} shows the region of the dynamic
    spectrum prior to and during the flaring episode, highlighting the oscillating flux level at a given frequency. The two vertical dashed lines in the dynamic spectrum denote the times of the images in Figure \ref{fig:clean_im_freq}.
    
\section{Statistics of Source Sizes}

\begin{figure}[htb!]
        \centering
        \includegraphics[width=1\textwidth]{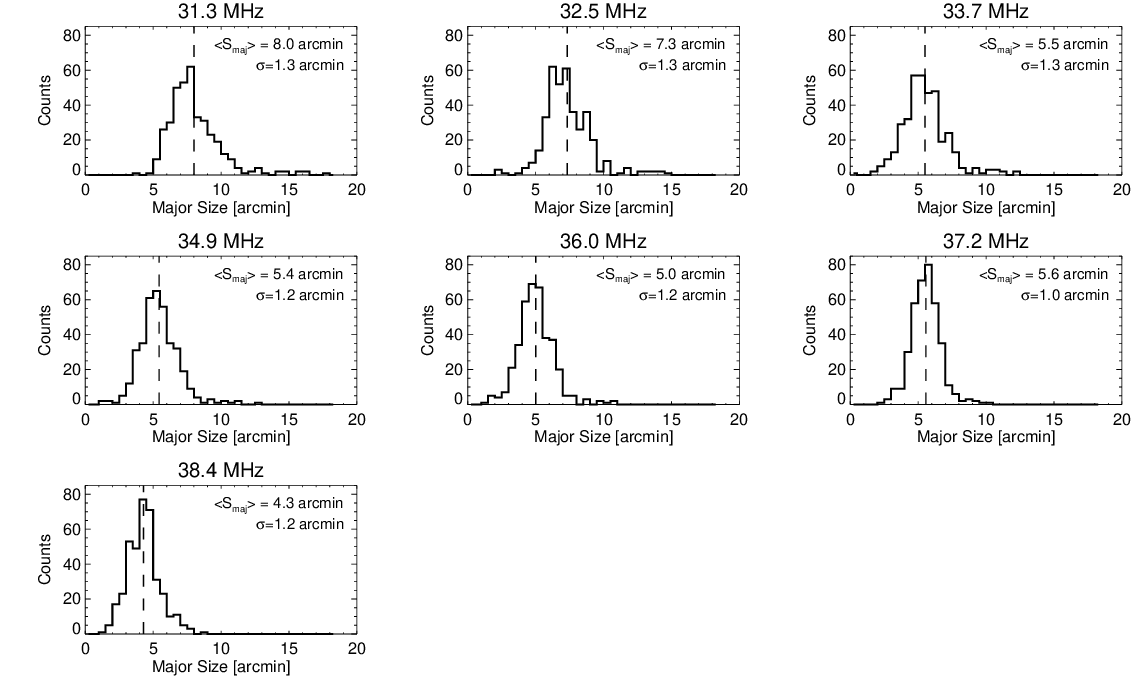}
        \caption{Noise storm sizes over 100 minutes. Each histogram
        represents one frequency channel with an $\sim1$~MHz
        bandwidth, and forms the 2D histogram shown in Figure
        \ref{fig:typeI_sizes}.}
        \label{fig:sizes_1Dhist}
    \end{figure}

    Figure \ref{fig:sizes_1Dhist} shows 1D histograms of the
    cleaned and beam-corrected source sizes over the 105~minute observation window,
    where the time resolution is averaged to 15.7~s, and the frequency
    is averaged to $\sim1$~MHz bandwidths. Each image is fit with a 2D Gaussian and
    the size estimated as described in section \ref{sec:data_methods}.
    Each histogram is comprised of 380 measurements.
    The vertical dashed lines in each panel mark the average size determined from each
    distribution as $\langle S_\mathrm{maj}\rangle=\left(\sum_i{N_iS_i}\right)\ \big/\sum_i N_i$ where $S_i$
    is the bin centered major size and $N_i$ is the number of measurements in that bin.
    The standard deviation $\sigma$ is given by
    $\sigma = \sqrt{\sum_i N_i \left(S_i - \langle S_{\mathrm{maj}} \rangle \right)^2 \big/ \sum_i N_i}$.

\bibliographystyle{aasjournal}
\bibliography{refs}

\end{document}